\documentclass{article} 
\usepackage{iclr2027_conference,times}

\usepackage{amsmath,amsfonts,bm}

\def\eqref#1{equation~\ref{#1}}

\def\1{\bm{1}}

\DeclareMathAlphabet{\mathsfit}{\encodingdefault}{\sfdefault}{m}{sl}
\SetMathAlphabet{\mathsfit}{bold}{\encodingdefault}{\sfdefault}{bx}{n}

\iclrfinalcopy

\usepackage[pagebackref=true,colorlinks,linkcolor=blue,citecolor=brown]{hyperref}
\usepackage{url}
\usepackage{circledsteps}
\usepackage{graphicx}
\usepackage{booktabs}
\usepackage{caption}

\usepackage{amsmath,amssymb}
\usepackage{wrapfig}
\usepackage{etoc}
\etocsettagdepth{mainpaper}{none}
\etocsettagdepth{appendix}{subsubsection}
\usepackage{algorithm}
\usepackage{algpseudocode}
\usepackage{xcolor}
\usepackage{amsmath,amssymb}
\usepackage{xcolor}
\usepackage[most]{tcolorbox}
\usepackage{tabularx,booktabs}
\definecolor{mblueL}{HTML}{C8D6D7}\definecolor{mblueD}{HTML}{87A3A7}
\definecolor{mredL}{HTML}{DCCAC7}\definecolor{mredD}{HTML}{B68E89}
\definecolor{mpurpleL}{HTML}{D2CCD9}\definecolor{mpurpleD}{HTML}{9F90AD}
\definecolor{mgreenL}{HTML}{C3CEBD}\definecolor{mgreenD}{HTML}{8F9D86}
\definecolor{mtext}{HTML}{30343A}
\newtcblisting[auto counter,number within=section]{promptbox}[4][mblueL]{%
  listing only, breakable, enhanced,
  colback=#1!45!white, colframe=#2, coltitle=white, colbacktitle=#2,
  boxrule=0.5pt, arc=2.5pt, left=4pt, right=4pt, top=3pt, bottom=3pt,
  fonttitle=\small\bfseries, title={Prompt~\thetcbcounter: #3}, label={#4},
  listing options={basicstyle=\ttfamily\scriptsize\color{mtext}, breaklines=true, breakatwhitespace=false,
                   columns=fullflexible, keepspaces=true, showstringspaces=false,
                   breakindent=0pt, postbreak={}, lineskip=-0.5pt}}

\title{Learning Self-Orchestration in \\Decentralized Agent Swarms}

\author{%
  Guowei Xu$^{*}$\quad
  Zhenting Qi$^{*}$ \quad
  Amani Kiruga \quad
  Himabindu Lakkaraju \quad
  Yilun Du \\[6pt]
  \textnormal{Harvard University \quad $^{*}$Equal contribution}
}

\newcommand{\ours}{\texttt{MPT}}

\begin{document}

\maketitle
\etocdepthtag.toc{mainpaper}

\begin{abstract}
The capabilities of an agent swarm depend on both its individual members and how they coordinate. Yet existing multi-agent systems often train agents within fixed orchestration structures, separating the learning of individual capabilities from the organization of collective behavior. We introduce market-based post-training (\ours{}), a framework for jointly learning action policies and self-orchestration in decentralized agent swarms. Agents express their willingness to participate through bids, which determine who acts and how collective outcomes provide local training credit. Learning both action and bidding behavior allows individual capabilities and patterns of participation to adapt together without a central coordinating agent. We evaluate \ours{} on embodied crafting, robot navigation, and deep research. Results show gains over monolithic post-training and confirm that agent participation indeed evolves with learning. With the same post-trained action policies, learned self-orchestration outperforms alternative coordination strategies. Ablations show that both action and bidding benefit from learning, and that learned bids provide more effective action-training signals. Beyond the original training objectives, trained swarms adapt faster to related tasks and achieve better performance with larger populations at inference time. Larger populations also suffer smaller performance losses under agent dropout, demonstrating that learned self-orchestration supports reuse, scaling, and robustness in decentralized agent swarms. More details are available on our project page at \url{https://guoweixu.com/mpt/}.

\end{abstract}

\begin{figure}[h]
\centering
\includegraphics[width=0.84\linewidth]{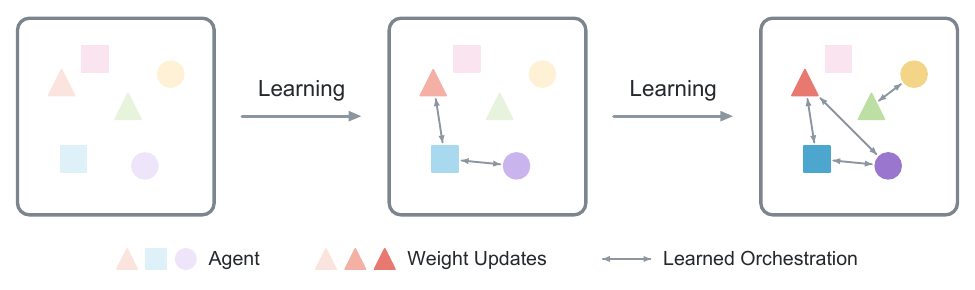}
\caption{\textbf{Learning self-orchestration in decentralized agent swarms.} The three panels illustrate successive stages of training within the same agent population. Each shape represents an agent, darker colors indicate progressive parameter updates, and bidirectional arrows depict coordination patterns. Through experience, agents jointly learn how to act and when to participate through bidding. As individual capabilities improve, patterns of participation adapt in turn.}
\label{fig:teaser}
\end{figure}

\section{Introduction}
\label{sec:intro}

Building general intelligence has long been a central ambition of AI research. But where does intelligence reside? A prevailing view in modern AI is model-centric: intelligence is contained in the parameters of a model, and progress comes from increasing model size or improving those parameters through training.
In this paper, we explore a different possibility. Intelligent behavior can also emerge from interactions among many simpler agents, giving rise to collective capabilities that no single agent possesses. This perspective echoes Minsky's Society of Mind~\citep{som}, which explains intelligence through the organization and interaction of many individually limited components.

Large language model (LLM) agents now exhibit strong individual reasoning capabilities, yet agent swarms remain limited in their ability to organize their own participation and actions toward a shared objective. Existing multi-agent systems typically place models within predefined interaction structures: a manager delegates to workers, a verifier checks a generator, or agents communicate according to a prescribed workflow~\citep{metagpt,debate,camel,autogen}. Even when the participating agents are trainable, the orchestration policy that determines which agent acts next often remains fixed throughout training~\citep{drmas}.

This separation overlooks the coupling between individual capabilities and swarm orchestration. Which agent acts in a particular state shapes not only the swarm's trajectory, but also which agents receive experience and learning signals. As agents improve, effective patterns of participation can change in turn. An agent that should act early in training may become unnecessary later; an action that was previously unhelpful may become valuable after downstream agents improve. Training individual agents under fixed orchestration therefore optimizes only one side of this coupled process. We instead ask:

\begin{center}
\textit{Can decentralized agent swarms jointly learn to act and self-orchestrate from experience?}
\end{center}

Learning self-orchestration in decentralized agent swarms poses a joint orchestration and credit-assignment problem. At each turn, agents must determine who should act. When task feedback becomes available, the swarm must also determine how to train the agents whose actions shaped the trajectory. These decisions are coupled: participation determines which agents acquire experience, while learning changes which and how agents should participate in the future. 

We therefore introduce market-based post-training (\ours{}), a framework that enables decentralized agent swarms to learn both how to act and how to self-organize their participation. Following \cite{eom}, agents compete for opportunities to act through bids that reflect their willingness to participate in the current decision. An economic incentive structure connects these participation decisions with credit assignment, using collective outcomes to improve both action and bidding behavior. As agents become more capable, the swarm learns when to rely on each member, allowing individual capabilities and self-orchestration to evolve together without a central coordinating agent.

We evaluate \ours{} on embodied crafting, robot navigation, and deep research. Comparisons on these benchmarks show higher training rewards than monolithic post-training, while further analysis reveals that agent participation evolves during learning. To isolate the contribution of self-orchestration, we compare coordination strategies using the same post-trained action policies. On Iron Pickaxe, \ours{} achieves a 63.4\% relative improvement in mean episode reward over the strongest alternative, Graph-of-Agents. We further show that both bidding and action benefit from learning, with their joint use yielding the strongest performance. A credit-assignment ablation shows that learned bids also provide more effective action-training signals.

We also examine the reusability, scalability, and robustness of \ours{}. Trained swarms adapt more rapidly than the single-agent baseline to new crafting objectives. Increasing the inference-time population improves pass@3 across crafting milestones without additional training, supporting the scalability of learned self-orchestration. Larger populations also better tolerate occasional agent unavailability. These results suggest that learned self-orchestration supports both reuse across objectives and flexible deployment with varying agent populations.

Our contributions are threefold:

(1) We introduce \ours{}, a framework for jointly learning  action policies and self-orchestration in decentralized agent swarms, allowing patterns of participation to adapt as agents improve.

 (2) We evaluate \ours{} on embodied crafting, robot navigation, and deep research, demonstrating gains over monolithic post-training. We further show the benefits of learned self-orchestration, learning both bidding and action, and using learned bids for credit assignment.

 (3) We demonstrate the reusability, scalability, and robustness of trained agent swarms: they adapt faster to new tasks, while larger inference-time populations improve performance and robustness.

\clearpage
\section{Method}
\label{sec:method}

We aim to jointly learn specialized action policies and self-orchestration in decentralized agent swarms. Each agent uses its local context to decide how strongly to participate through bidding. Bids determine both which agent acts and how task rewards are allocated across participating agents for learning. We first describe this orchestration mechanism (\S\ref{sec:method-market}), then explain how agents learn to act from the resulting credit (\S\ref{sec:method-act}) and learn to bid through counterfactual comparisons (\S\ref{sec:method-bid}).

\subsection{Self-Orchestration through Bidding}
\label{sec:method-market}

\begin{figure}[t]
\centering
\includegraphics[width=\linewidth]{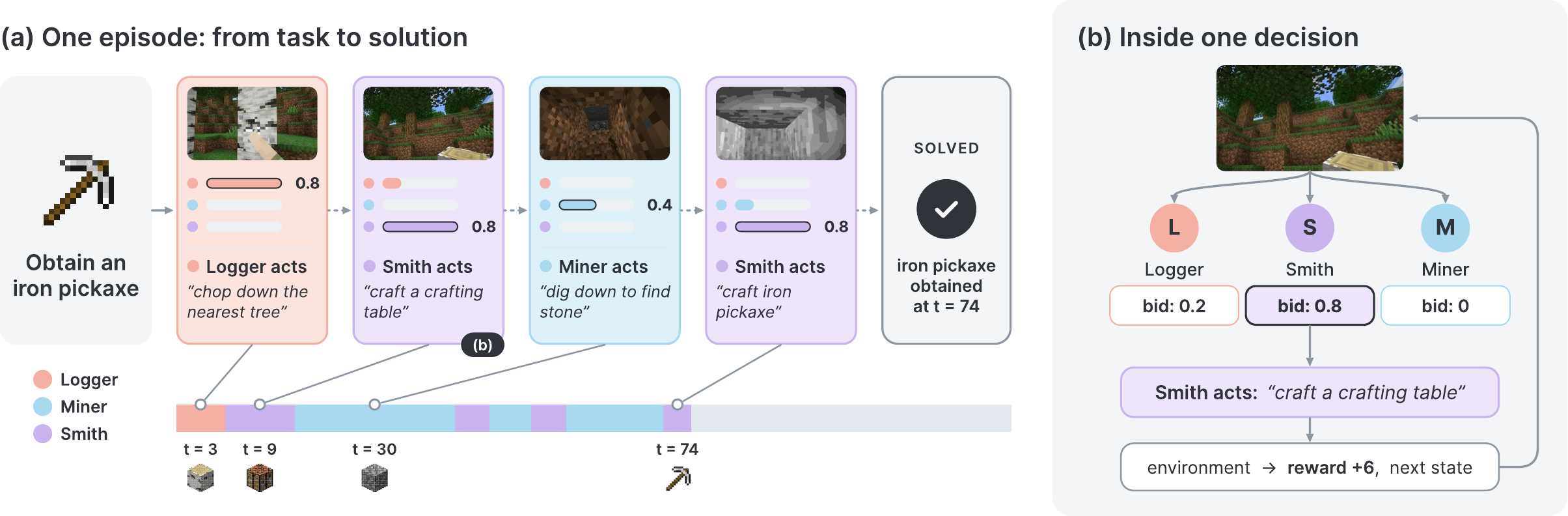}
\caption{\textbf{Self-orchestration through bidding in embodied crafting.} (a) Selected turns from an Iron Pickaxe episode show the Logger, Miner, and Smith participating as resource gathering and crafting requirements change. (b) At each decision, agents submit bids, and the highest bidder acts. Here, the Smith wins with a bid of 0.8 and crafts a crafting table; the environment then returns a reward and the next state.}
\label{fig:case}
\end{figure}

We consider episodic tasks with verified rewards. An episode presents a task $x$ and unfolds over turns $t = 1, \dots, T$, with one agent acting at each turn and a verifier assigning a task reward $R$. The swarm consists of $N$ agents with separate model parameters. Each agent $i$ uses a policy $\pi_i$ to generate both action and bidding. 

Following~\citet{eom}, we adopt an auction mechanism for decentralized turn allocation. At each turn, agents independently submit bids $b_i$ reflecting their willingness to participate in the current decision. The agent with the highest bid is selected to act, and its bid is denoted by $b_t$. This allows participation to depend on the agents' local assessments without a central dispatcher.  Tied bids are broken uniformly at random. If all agents bid zero, no agent acts and the turn is skipped.

Bids also connect participation with credit assignment. Specifically, the selected bid $b_t$ determines the fraction of accumulated value at turn $t$ that is passed backward to the preceding acting agent. Thus, a larger bid increases an agent's opportunity to act while committing a larger share of the resulting value to earlier participation. Figure~\ref{fig:case} illustrates how agents bid and act based on the current task state.

\subsection{Learning to Act}
\label{sec:method-act}

\begin{figure}[h]
\centering
\includegraphics[width=\linewidth]{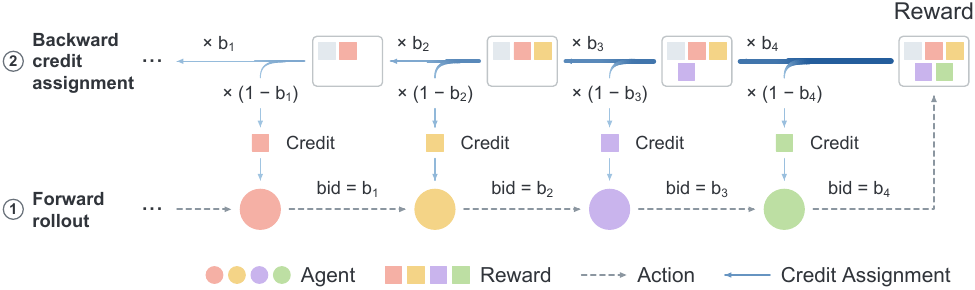}
\caption{\textbf{Learning to act through bidding-based credit assignment.} Agents act sequentially during the forward rollout (\Circled{1}). The resulting reward is then propagated backward (\Circled{2}): at turn $t$, the acting agent retains a fraction $1-b_t$ of the accumulated value as its credit and passes the remaining fraction $b_t$ to the preceding participant. Colored blocks associate credit with its recipient.}
\label{fig:act}
\end{figure}

At the end of each episode, we assign local training credit by propagating rewards backward through the sequence of acting agents. Let $\hat r_t$ denote the environment reward obtained at turn $t$, with the terminal verified reward included in $\hat r_T$. We define the accumulated value $v_t$ and the local training credit $\tilde r_t$ as
\begin{equation}
v_t = \hat r_t + b_{t+1} v_{t+1},
\qquad
\tilde r_t = (1-b_t)v_t,
\qquad
v_{T+1}=0.
\label{eq:chain}
\end{equation}

The agent acting at turn $t$ receives credit from both its immediate reward and the value passed backward from the next turn. It retains a fraction $1-b_t$ for its own action update and passes the remaining fraction backward. The share passed backward by the first acting agent is discarded. A reward obtained at turn $t+k$ contributes to $v_t$ with weight $\prod_{j=t+1}^{t+k} b_j$. This allows downstream outcomes to provide learning signals for earlier actions, following the principle of a bucket brigade~\citep{holland,baum}. Figure~\ref{fig:act} illustrates this backward credit assignment.

Each agent updates its own parameters using only the action tokens from turns in which it acted. We use REINFORCE~\citep{reinforce} with a per-agent mean baseline:
\begin{equation}
\nabla_{\theta_i} J
=
\mathbb{E}_{t \in \mathcal{W}_i}
\left[
(\tilde r_t-\bar r_i)
\nabla_{\theta_i}\log \pi_i(a_t \mid h_{i,t})
\right],
\label{eq:reinforce}
\end{equation}
where $\mathcal{W}_i$ contains the turns assigned to agent $i$, and $\bar r_i$ is its mean training credit over those turns in the batch.

\subsection{Learning to Self-Orchestrate}
\label{sec:method-bid}

\begin{figure}[h]
\centering
\includegraphics[width=0.85\linewidth]{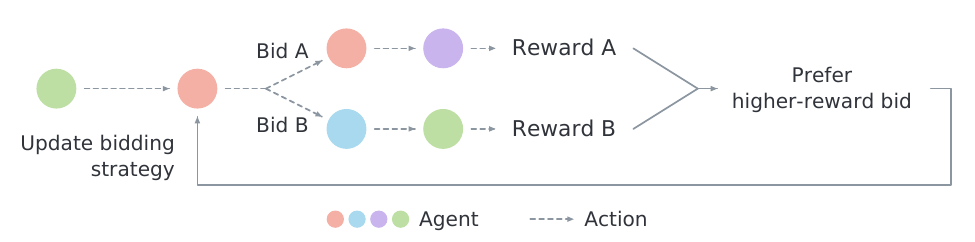}
\caption{\textbf{Learning to self-orchestrate through counterfactual comparisons.} Starting from the same state, two alternative bids from the same agent induce different turn allocations. The alternative branch is continued from the restored state, and both branches are evaluated, producing returns that determine which bid is preferred. Preference learning then increases the relative likelihood of the bid associated with the better outcome, teaching the agent when to participate.}
\label{fig:bid}
\end{figure}

Learning to self-orchestrate requires feedback about participation: would the trajectory have improved if another agent had acted at the same turn? We obtain this feedback by comparing counterfactual rollouts from the same state, as illustrated in Figure~\ref{fig:bid}.

We define a turn as contested if the runner-up submits a nonzero bid. The
original branch is the collected trajectory, in which the selected agent
acts at the contested turn. To construct the alternative branch, we restore
the environment and agent states at the contested turn in a replica and
resample the selected agent's bid up to three times until it falls below the runner-up's bid, so that the runner-up acts instead. If no such bid is obtained, the contested turn is discarded from bidding training. The
alternative branch is then continued until the episode terminates. Each branch is scored by its sum of environment rewards from the contested turn onward. The bidding response associated with the higher-scoring branch becomes the preferred response $y_w$, and the other becomes the rejected response $y_l$. Pairs with equal branch returns are discarded.
We train the agent on these response pairs using a Bradley--Terry preference objective~\citep{prefer}:
\begin{equation}
\mathcal{L}_{\mathrm{bid}}
=
-\log \sigma(\log \pi_i(y_w \mid h_{i,t})
-
\log \pi_i(y_l \mid h_{i,t})),
\label{eq:bidpref}
\end{equation}
where $\sigma$ is the logistic function. Minimizing this loss increases the relative likelihood of the bid associated with the better outcome. 
We alternate between action and bidding learning during training, using four action-training steps followed by one bidding-training step. 
Both updates optimize the same model parameters within each agent. This allows action competence and participation to co-adapt as learning changes the capabilities of the swarm.

\section{Experiments}
\label{sec:experiments}

\subsection{Setup}
\label{sec:exp_setup}

We evaluate \ours{} on decision-making and reasoning tasks spanning embodied crafting, robot navigation, and deep research. Each setting uses three agents with complementary roles.

\textbf{Benchmarks and Models.}
(1) \emph{Embodied crafting}: three Qwen3-VL-4B-Instruct agents (Logger, Miner, and Smith) specialize in gathering wood, mining resources, and crafting. Starting from an empty inventory in Minecraft MineStudio~\citep{minestudio}, they must progress through the crafting chain to obtain an iron pickaxe. Agents receive RGB observations and textual state information, including inventory and recent interaction history. The selected agent issues a language instruction to a frozen STEVE-1 controller~\citep{steve1}, which executes it for 40 game frames.

(2) \emph{Robot navigation}: three NaVILA-8B agents~\citep{navila} perform continuous vision-and-language navigation on VLN-CE R2R in Habitat~\citep{vlnce,r2r,habitat}. Given a route instruction and an eight-frame window of egocentric observations, the agents coordinate through partitioned action spaces: Turner controls rotations, Mover controls forward movement, and Stopper decides when to stop. Episodes contain at most 64 decisions.

(3) \emph{Deep research}: three \texttt{gemma-4-E2B-it} agents~\citep{gemma4} answer questions by browsing a searchable document corpus. Searcher, Opener, and Finder specialize in \texttt{browser.search}, \texttt{browser.open}, and \texttt{browser.find}; all agents share the interaction history and may submit a final answer. We use OpenResearcher~\citep{openresearcher} for training. Episodes contain at most 64 decisions and receive a binary terminal reward based on answer correctness, assessed against the reference answer by a \texttt{qwen3.5-flash} judge.

\textbf{Training Setup.}
All agents have independent model parameters and undergo a supervised fine-tuning (SFT) cold start before post-training. Each agent in \ours{} is fine-tuned on demonstrations corresponding to its role. We compare \ours{} against a monolithic REINFORCE baseline with a discount factor of $\gamma=0.95$. During the SFT cold start, the baseline is trained on the combined demonstrations from all roles, with total SFT training compute matched between the two settings. Detailed task settings and hyperparameters are provided in Appendix~\ref{app:task}. 

\subsection{Main Results}
\label{sec:exp_main}

\begin{figure}[h]
\centering
\includegraphics[width=\linewidth]{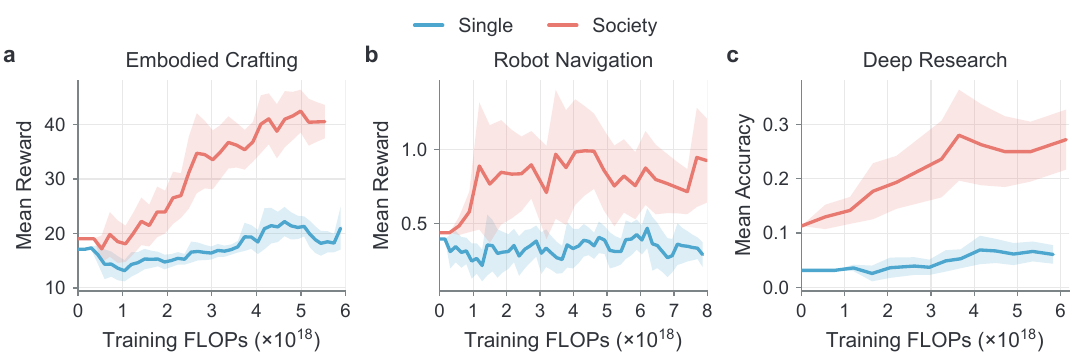}
\caption{\textbf{\ours{} improves performance at matched training compute.} Training performance versus training FLOPs for monolithic RL (Single) and \ours{} (Society). The panels report mean episode reward for embodied crafting and robot navigation, and mean accuracy for deep research. Training FLOPs include all compute used for learning.}
\label{fig:main}
\end{figure}

Figure~\ref{fig:main} compares \ours{} with monolithic RL under matched training FLOP budgets across three tasks. All methods are compared under matched training FLOPs, which count all LLM computation during training, including bid generation by all $N$ agents, action generation, counterfactual rollouts, and parameter updates. In embodied crafting, \ours{} achieves a final mean episode reward of approximately 41, compared with 20 for the monolithic baseline. In robot navigation, \ours{} maintains substantially higher rewards after the initial training phase, ending at approximately 0.9 versus 0.3 for the baseline. In deep research, \ours{} improves mean accuracy to approximately 27\%, whereas the monolithic baseline reaches about 6\%. Qualitative case studies are provided in Appendix~\ref{app:vis}.

\noindent
\begin{minipage}[c]{0.50\linewidth}
Table~\ref{tab:final_eval} reports results on held-out sets disjoint from training: 32 unseen world seeds for embodied crafting, 88 R2R-CE val-unseen episodes for robot navigation, and 50 held-out OpenResearcher questions for deep research. Each instance is evaluated three times with the training settings. \ours{} outperforms monolithic RL on all three tasks.
\end{minipage}\hfill
\begin{minipage}[c]{0.48\linewidth}
\centering
\captionsetup{type=table,font=small,skip=4pt}
\caption{\textbf{Held-out evaluation.} Mean $\pm$ SE over instances, each averaged over three runs.}
\label{tab:final_eval}
\small
\setlength{\tabcolsep}{3pt}
\renewcommand{\arraystretch}{1.1}
\begin{tabular}{@{}lccc@{}}
\toprule
Method & \shortstack{Embodied\\Crafting}
       & \shortstack{Robot\\Navigation}
       & \shortstack{Deep\\Research} \\
\midrule
RL & $20.3 \pm 4.1$ & $0.52 \pm 0.14$ & $9.0 \pm 2.1\%$ \\
\ours{} & $\mathbf{41.5 \pm 5.9}$
        & $\mathbf{1.23 \pm 0.18}$
        & $\mathbf{21.0 \pm  4.1\%}$ \\
\bottomrule
\end{tabular}
\end{minipage}
\par

\subsection{Analysis}
\label{sec:exp_abl}

\subsubsection{Self-Orchestration Evolves during Training}
\label{sec:exp_orchestration_learning}

\begin{figure}[h]
\centering
\includegraphics[width=\linewidth]{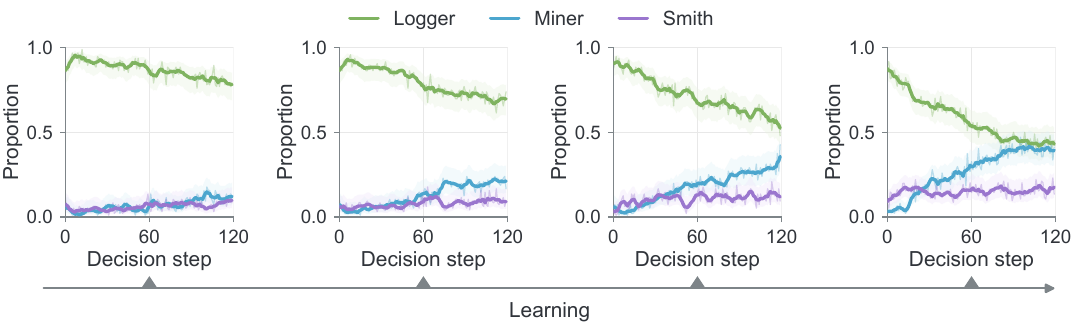}
\caption{\textbf{Self-orchestration evolves during training.} Participation proportions of the Logger, Miner, and Smith on Iron Pickaxe. From left to right: after SFT and before RL, two intermediate RL checkpoints, and after RL. As training progresses, participation increasingly shifts toward the Miner and Smith at later decision steps.}
\label{fig:stats}
\end{figure}

To examine how self-orchestration develops through learning, we track agent participation on the Iron Pickaxe task across four checkpoints: after SFT but before RL, at two intermediate stages of RL, and after RL. Figure~\ref{fig:stats} plots the participation proportions of the Logger, Miner, and Smith against decision step.

Before RL, the Logger dominates throughout the episode, while the Miner and Smith participate infrequently. As training progresses, this pattern changes: the Logger remains predominant at early steps, but its share decreases later as the Miner and Smith become more involved. This shift becomes more pronounced after RL, with the Miner approaching the Logger's participation level toward the end of the episode. At decision step 120, the Logger's participation share decreases from approximately 80\% before RL to 45\% after RL, while the Miner's share increases from approximately 10\% to 40\%. The Smith's share also increases from approximately 10\% to 15\%. These observations provide behavioral evidence that the swarm adapts its allocation of acting turns during training. 

\subsubsection{Comparison between Self-Orchestration and Other Coordination Strategies}
\label{sec:exp_orchestration}

We evaluate whether learned self-orchestration coordinates agents more effectively than alternative inference-time strategies. We keep the post-trained action policies fixed and compare our bidding mechanism with Round-Robin~\citep{autogen}, AgentNet~\citep{agentnet}, and Graph-of-Agents (GoA)~\citep{goa}. All configurations are evaluated on the Iron Pickaxe crafting task over the same 32 seeds.

As shown in Table~\ref{tab:orchestration} (left), \ours{} substantially outperforms all three baselines. Specifically, \ours{} achieves a mean episode reward of $41.5$, compared with $25.4$ for GoA, the strongest alternative coordination strategy, which corresponds to a 63.4\% relative improvement. These gains demonstrate that learned self-orchestration coordinates agents more effectively.

\subsubsection{Both Bidding and Action Benefit from Learning}
\label{sec:both_learn}

In this section, we show that training effectively improves both the action policy and self-orchestration. To verify this, we examine their contributions through four inference-time configurations, independently using either the initial or post-trained checkpoints for each component. We evaluate each configuration over the same 32 seeds and report mean episode reward in Table~\ref{tab:both_learn}.

As shown in the table, using post-trained checkpoints for both components substantially outperforms all other configurations, whereas using a post-trained checkpoint for only bidding or only action yields smaller improvements over the initial baseline. Using post-trained checkpoints for both components yields a mean reward of 41.5, compared with 21.5 for trained bidding with initial actions and 27.2 for initial bidding with trained actions. The fully post-trained configuration achieves 2.69 times the reward of the initial configuration, which obtains 15.4. These results indicate that strong action and bidding policies are both essential for effective self-orchestration.

\begin{table}[h]
\centering
\caption{\textbf{Learned self-orchestration improves coordination.} Episode reward (mean $\pm$ SE) on Iron Pickaxe over 32 seeds. Left: comparison with alternative coordination strategies using the same  policies. Right: contributions of learning bidding and action.}
\label{tab:orchestration}
\label{tab:both_learn}

\begin{minipage}[t]{0.51\linewidth}
\begin{tabular*}{\linewidth}{@{\extracolsep{\fill}}lc@{}}
\toprule
Strategy & Mean Reward \\
\midrule
Round-robin~\citep{autogen} & $23.8 \pm 4.2$ \\
AgentNet~\citep{agentnet} & $20.9 \pm 4.2$ \\
GoA~\citep{goa} & $25.4 \pm 4.6$ \\
\ours{} & $\mathbf{41.5 \pm 5.9}$ \\
\bottomrule
\end{tabular*}
\end{minipage}\hfill
\begin{minipage}[t]{0.47\linewidth}
\begin{tabular*}{\linewidth}{@{\extracolsep{\fill}}llc@{}}
\toprule
Bidding & Action & Mean Reward \\
\midrule
Initial & Initial & $15.4 \pm 4.1$ \\
Trained & Initial & $21.5 \pm 4.7$ \\
Initial & Trained & $27.2 \pm 4.8$ \\
Trained & Trained & $\mathbf{41.5 \pm 5.9}$ \\
\bottomrule
\end{tabular*}
\end{minipage}
\end{table}

\subsubsection{Isolating the Contributions of Learned Self-Orchestration}
\label{sec:cre}

\noindent
\begin{minipage}[t]{0.64\linewidth}
\vspace{0pt}
MPT differs from monolithic RL in several respects. To isolate the contribution of learned self-orchestration, we conduct two ablations on the Iron Pickaxe task. Both ablations share the same agent roles, SFT initialization, and action-training procedure as \ours{}, and each modifies a single component.

\textbf{Round-robin.} Turns are allocated cyclically among the three agents, and policies are trained on the reward with a discount factor of $\gamma = 0.95$. This ablation removes bidding entirely.

\textbf{Random credit.} We replace the bids used for credit assignment with random values sampled from $\mathcal{U}(0,1)$, while turn allocation and bidding optimization remain unchanged. This ablation tests whether learned bids provide useful credit-assignment signals.

As shown in Figure~\ref{fig:abl}, at approximately $5 \times 10^{18}$ training FLOPs, \ours{} reaches a mean episode reward of about 42, compared with 15 for Round-robin, and 31 for random credit. Round-robin improves little over training. The gap to random credit shows that learned bids also provide more effective credit assignment.

\end{minipage}\hfill
\begin{minipage}[t]{0.33\linewidth}
\vspace{0pt}
\captionsetup{type=figure}
\centering
\includegraphics[width=\linewidth]{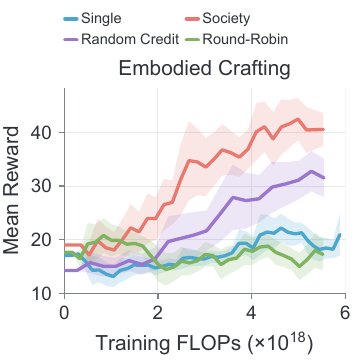}
\caption{\textbf{Ablation studies on Iron Pickaxe.} Curves are plotted against training FLOPs.}
\label{fig:abl}
\end{minipage}
\par

\subsection{Reusability, Scalability, and Robustness}

\paragraph{Reusing learned self-orchestration.}

\begin{figure}[h]
\centering
\includegraphics[width=0.67\linewidth]{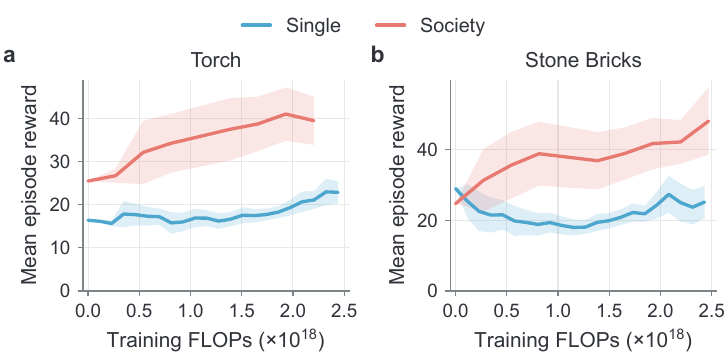}
\caption{\textbf{Trained swarms adapt faster to new objectives.} Adaptation to (a) Torch and (b) Stone Bricks. Monolithic RL (Single) and \ours{} (Society) start from their respective Iron Pickaxe checkpoints and continue training separately. Curves show mean episode reward against training FLOPs.}
\label{fig:continual}
\end{figure}

To support a growing range of tasks, agent swarms need to build on previously learned skills and coordination. We investigate whether learned self-orchestration can be reused to accelerate adaptation to new objectives. We evaluate this through continual learning: starting from their Iron Pickaxe checkpoints, we continue training \ours{} and monolithic RL separately on two additional crafting tasks, Torch and Stone Bricks. Figure~\ref{fig:continual} reports mean episode reward against the training FLOPs.

The society adapts more rapidly and achieves higher rewards at comparable training FLOP budgets on both tasks. At approximately $10^{18}$ adaptation FLOPs, the society achieves mean episode rewards of about 35 on Torch and 38 on Stone Bricks, compared with 17 and 19 for the monolithic baseline.  These results support reusing the trained agent population for efficient adaptation to new objectives.

\begin{figure}[h]
\centering
\includegraphics[width=\linewidth]{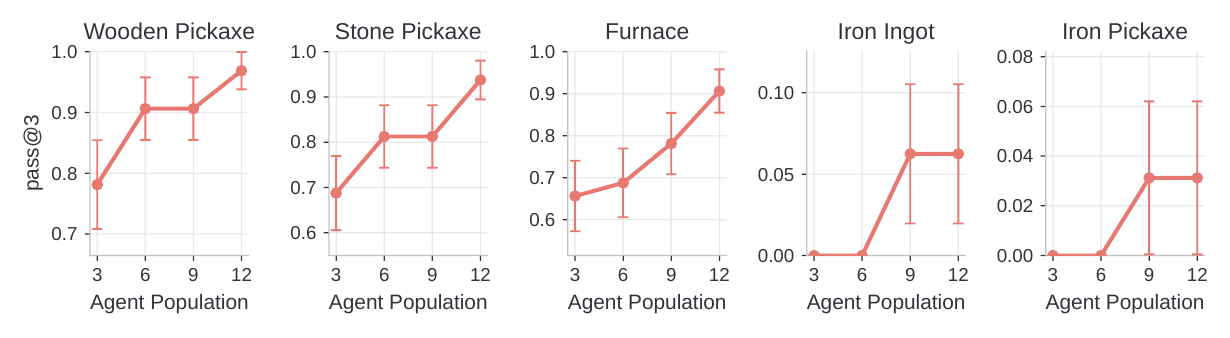}
\caption{\textbf{Larger swarms improve inference-time performance.} Pass@3 across five crafting milestones as the agent population increases from 3 to 12.}
\label{fig:its}
\end{figure}

\paragraph{Inference-time scaling and robustness.}
Decentralized agent swarms should scale at inference time. We investigate whether learned self-orchestration enables such scaling through larger agent populations. We vary the population size across 3, 6, 9, and 12 agents by using one to four copies of each trained agent. We conduct experiments on the Iron Pickaxe crafting task over 32 seeds, with three runs per seed, and evaluate pass@3 on five subtasks: Wooden Pickaxe, Stone Pickaxe, Furnace, Iron Ingot, and Iron Pickaxe. Figure~\ref{fig:its} shows that larger populations improve performance on all subtasks. Increasing the population from 3 to 12 agents raises pass@3 from approximately 78\% to 97\% on Wooden Pickaxe, from 69\% to 94\% on Stone Pickaxe, and from 66\% to 91\% on Furnace. These results indicate that self-orchestration can effectively utilize additional agents to support inference-time scaling.

\noindent
\begin{minipage}[t]{0.6\linewidth}
\vspace{0pt}
Another advantage of decentralized agent swarms is that they do not rely on a central agent. This can limit the impact of occasional agent unavailability, which becomes increasingly likely as the population grows. To evaluate this, we introduce agent dropout at inference time and measure the reward under dropout. We conduct experiments on the Iron Pickaxe task over 32 seeds, averaging three runs per seed. Figure~\ref{fig:robust} shows that larger populations can preserve more capability even under severe dropout. At a dropout probability of $p=0.9$, the 12-agent population maintains a mean episode reward of approximately 33, compared with 19 for the 3-agent population. These results suggest that additional agents provide redundancy that helps the swarm maintain its capabilities when members become unavailable.
\end{minipage}\hfill
\begin{minipage}[t]{0.33\linewidth}
\vspace{0pt}
\captionsetup{type=figure}
\centering
\includegraphics[width=\linewidth]{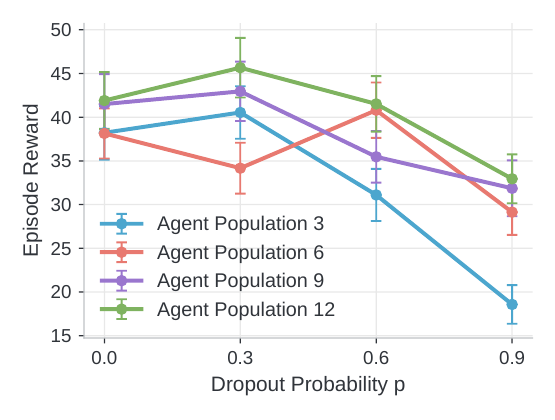}
\caption{\textbf{Larger swarms better tolerate agent dropout.} Mean episode reward on Iron Pickaxe across dropout rates.}
\label{fig:robust}
\end{minipage}
\par

\section{Related Work}
\label{sec:related}

\paragraph{LLM-based multi-agent systems.}
LLM-based multi-agent systems combine role specialization, communication, and memory to support collective behavior. Generative Agents studies emergent social interactions among agents with memory and reflection~\citep{generative_agents}, while CAMEL explores cooperation through role-playing~\citep{camel}. MetaGPT and ChatDev organize software development through structured workflows and specialized roles~\citep{metagpt,chatdev}; AutoGen and AgentVerse provide frameworks for programmable conversations and collaborative agent teams~\citep{autogen,agentverse}. Multiagent debate improves reasoning through iterative exchange and revision of proposed answers~\citep{debate}. Beyond manually specified workflows, GPTSwarm optimizes graph connectivity~\citep{gptswarm}, AgentNet adapts decentralized routing through experience and evolving connections~\citep{agentnet}, and Graph-of-Agents selects agents and refines responses through graph-based collaboration~\citep{goa}. 
However, these systems generally adapt interactions at inference time. \ours{} learns policies through weight updates, allowing agent capabilities and self-orchestration to improve together.

\paragraph{Multi-agent reinforcement learning.}
Learning decentralized policies from collective feedback is a central problem in multi-agent reinforcement learning. MADDPG and MAPPO use centralized training with decentralized execution~\citep{maddpg,mappo}, while VDN and QMIX factorize team value functions into agent-level utilities~\citep{vdn,qmix}. COMA addresses credit assignment through a counterfactual baseline that marginalizes one agent's action while holding the others fixed~\citep{coma}. Coordination is also learned explicitly: RIAL and DIAL learn communication protocols~\citep{rial_dial}, CommNet jointly learns continuous communication and control~\citep{commnet}, and IC3Net learns when agents should communicate~\citep{ic3net}.  Recent work extends multi-agent training to LLMs: Multiagent Finetuning updates specialized models using their own interaction data~\citep{multiagent_ft}; Dr.\ MAS stabilizes reinforcement learning through agent-wise advantage normalization~\citep{drmas}; and CoLLM studies centralized and decentralized critics for training decentralized LLM collaborators~\citep{collmdc}. 
However, these methods do not connect the choice of the next agent with how downstream reward is distributed across agents. \ours{} couples turn allocation and credit assignment through learned bids, using task feedback to jointly improve participation and policies.

\paragraph{Economics in agents.}
Economic coordination offers a longstanding perspective on distributed intelligence. Hayek emphasizes how prices communicate dispersed knowledge~\citep{hayek}, while Minsky describes intelligence as arising from interactions among individually limited components~\citep{som}. The Contract Net Protocol allocates tasks through negotiation~\citep{contract_net}, and market-oriented programming uses  markets for decentralized resource allocation~\citep{market_programming}. Economic mechanisms also connect participation to learning: Holland's bucket brigade transfers credit along chains of active rules~\citep{holland}, and Baum's Hayek Machine develops economic credit assignment~\citep{baum1996,baum}. More recently, Economy of Minds uses auctions to coordinate LLM agents and guide the evolution of prompts and population with frozen model weights~\citep{eom}. SALE selects among agents' proposed strategies using cost--value scoring and shared auction memory~\citep{sale}. 
However, recent LLM-based economic systems primarily use feedback to  evolve prompts while keeping weights frozen. \ours{} uses feedback to train action and bidding policies, enabling agents to learn how to act and when to participate.
\section{Conclusion}
\label{sec:conclusion}

We introduced \ours{}, a framework for jointly learning individual action policies and self-orchestration in decentralized agent swarms. Learned bids connect participation decisions with local credit assignment, allowing agents to improve both how they act and when they participate without a central coordinating agent. Our experiments show benefits over monolithic post-training and other coordination strategies, while ablations demonstrate the value of learning both bidding and action policies and using learned bids for credit assignment. Trained swarms also adapt faster to related objectives, benefit from larger populations at inference time, and better tolerate agent unavailability as the population grows. These findings support treating self-orchestration as a learnable and reusable component of collective intelligence, opening a path toward decentralized agent swarms whose capabilities and organization improve together through experience.

\clearpage

\bibliography{iclr2027_conference}
\bibliographystyle{iclr2027_conference}

\clearpage
\appendix

\etocdepthtag.toc{appendix}
\etocsettocstyle{\section*{Appendix - Table of Contents}}{}
\tableofcontents
\clearpage

\section{Pseudo Code}
\label{app:algo}

We provide the full pseudocode for \ours{} below. Algorithm~\ref{alg:ours} describes alternating action and bidding updates, using credit assignment to train actions and counterfactual comparisons to train bidding.

\begin{algorithm}[h]
\caption{\ours{}: Learning to Act and Self-Orchestrate}
\label{alg:ours}
\begin{algorithmic}[1]
\small

\Require Agent policies $\{\pi_{\theta_i}\}_{i=1}^{N}$;
task distribution $\mathcal{D}$; training iterations $K$;
learning rates $\eta_{\mathrm{act}}, \eta_{\mathrm{bid}}$.

\For{$k=1,\ldots,K$}

    \Statex \textcolor{gray}{\# Collect on-policy trajectories}
    \State $\mathcal{B} \gets
    \Call{CollectRollouts}{\mathcal{D},\{\pi_{\theta_i}\}_{i=1}^{N}}$
    \Statex \textcolor{gray}{\quad Agents bid, and the highest bidder acts.}

    \If{$k \bmod 5 \neq 0$}

        \Statex \textcolor{gray}{\# Learn to act}
        \State $\mathcal{A}_i \gets \emptyset$
        for all $i=1,\ldots,N$

        \For{each trajectory $\tau \in \mathcal{B}$}
            \State $v_{T+1}\gets 0$; $b_{T+1}\gets 0$

            \For{$t=T,\ldots,1$}
                \State $v_t \gets \hat r_t+b_{t+1}v_{t+1}$
                \State $\tilde r_t \gets (1-b_t)v_t$
                \State $\mathcal{A}_{i_t} \gets
                \mathcal{A}_{i_t}\cup
                \{(h_{i_t,t},a_t,\tilde r_t)\}$
            \EndFor
        \EndFor

        \For{each agent $i$ with $\mathcal{A}_i\neq\emptyset$}
            \State $\bar r_i \gets
            \frac{1}{|\mathcal{A}_i|}
            \sum_{(h,a,c)\in\mathcal{A}_i}c$
            \State $\displaystyle
            \theta_i \gets \theta_i+
            \frac{\eta_{\mathrm{act}}}{|\mathcal{A}_i|}
            \sum_{(h,a,c)\in\mathcal{A}_i}
            (c-\bar r_i)\nabla_{\theta_i}
            \log\pi_{\theta_i}(a\mid h)$
        \EndFor

    \Else

        \Statex \textcolor{gray}{\# Learn to self-orchestrate}
        \State $\mathcal{P}_i \gets \emptyset$
        for all $i=1,\ldots,N$
        \State $\mathcal{C}\gets
        \Call{SampleContestedTurns}{\mathcal{B}}$

        \For{each $(\tau,t,i,h,y^{\mathrm{orig}})\in\mathcal{C}$}
            \State $R^{\mathrm{orig}} \gets \sum_{t'\ge t}\hat r_{t'}$ in $\tau$
            \State Resample $y^{\mathrm{alt}}$ from agent $i$
            up to 3 times until its bid lets the runner-up act
            \If{no such $y^{\mathrm{alt}}$ is obtained}
                \State \textbf{continue}
            \EndIf
            \State $R^{\mathrm{alt}}
            \gets
            \Call{AlternativeRollout}
            {\tau,t,y^{\mathrm{alt}}}$

            \If{$R^{\mathrm{orig}}\neq R^{\mathrm{alt}}$}
                \State $(y_w,y_l)\gets
                (y^{\mathrm{orig}},y^{\mathrm{alt}})$
                ordered by decreasing branch reward
                \State $\mathcal{P}_i\gets
                \mathcal{P}_i\cup\{(h,y_w,y_l)\}$
            \EndIf
        \EndFor

        \For{each agent $i$ with $\mathcal{P}_i\neq\emptyset$}
            \State $\displaystyle
            \mathcal{L}_i \gets
            -\frac{1}{|\mathcal{P}_i|}
            \sum_{(h,y_w,y_l)\in\mathcal{P}_i}
            \log\sigma\!\left(
            \log\pi_{\theta_i}(y_w\mid h)
            -\log\pi_{\theta_i}(y_l\mid h)
            \right)$
            \State $\theta_i\gets
            \theta_i-\eta_{\mathrm{bid}}
            \nabla_{\theta_i}\mathcal{L}_i$
        \EndFor

    \EndIf
\EndFor
\State \Return $\{\pi_{\theta_i}\}_{i=1}^{N}$
\end{algorithmic}
\end{algorithm}

\section{Detailed Experimental Setup}
\label{app:task}

\subsection{Embodied Crafting}
\label{app:task_ec}

\subsubsection{Task and Simulator}
The task is to obtain an iron pickaxe from an empty inventory in a randomly generated Minecraft world, which requires the tech tree: chop logs, craft planks, a crafting table and sticks, craft and equip a wooden pickaxe, mine cobblestone, craft a stone pickaxe and a furnace, mine iron ore, smelt iron ingots, and finally craft the iron pickaxe. The simulator is Minecraft through MineStudio~\citep{minestudio}. Each world is identified by a seed. All post-training runs draw from 1614 seeds.

\subsubsection{Reward}
The environment pays a staged, event-based reward defined over twelve tech-tree events $\mathcal{E}$, each with a per-occurrence reward $r_e$ and a per-episode cap $c_e$ on how many occurrences are paid: mine a log ($r_e{=}1$, $c_e{=}4$), craft planks (2, 3), craft a crafting table (3, 1), open the crafting table (3, 1), craft a stick (4, 1), craft a wooden pickaxe (8, 1), mine cobblestone (4, 3), craft a stone pickaxe (20, 1), craft a furnace (8, 1), mine iron ore (4, 3), smelt an iron ingot (10, 3), and craft the iron pickaxe (30, 1). Let $n_e(t)$ be the number of times event $e$ has occurred in the episode up to and including decision $t$. The reward paid at decision $t$ and the episode return are
\begin{equation}
r_t=\sum_{e\in\mathcal{E}} r_e\,\Big[\min\big(n_e(t),c_e\big)-\min\big(n_e(t-1),c_e\big)\Big],\qquad
R=\sum_{t=1}^{120} r_t\;\le\;\sum_{e\in\mathcal{E}} r_e\,c_e=140,
\label{eq:ec_reward}
\end{equation}
so every event pays a fixed amount for each of its first $c_e$ occurrences and nothing afterwards, and reward sums to 140.

\subsubsection{Episode, Observation and Decision}
An episode contains 120 decisions. At each decision the policy receives a $128\times128$ RGB screenshot and a text description: the task, crafting rules, current decision step, cumulative events, world status, a rolling history of the last six steps, the live inventory, and the outcome of the previous instruction. The policy answers with reasoning followed by a natural-language instruction, \texttt{<think>...</think><action>...</action>}. Instructions are executed by a frozen text-conditioned low-level controller, STEVE-1~\citep{steve1}, for 40 game frames. 

\subsubsection{Agent Design}
The population consists of three agents, \textsc{Logger}, \textsc{Miner} and \textsc{Smith}, each a separate full fine-tune of Qwen3-VL-4B-Instruct~\citep{qwen3vl}. 
The monolithic agent is the same base model fine-tuned on instructions spanning all three roles. 

Below, we present the example prompt for the monolithic agent and the example bidding and action prompts for each agent in the society.

\textbf{Single Agent.}

\begin{promptbox}[mblueL]{mblueD}{Single Agent}{box:ec_single}
# Screenshot
<image>

TASK: Obtain an iron pickaxe. You start with nothing. Chop trees for logs, craft planks, a crafting table and sticks, craft and equip a wooden pickaxe, dig down and mine cobblestone, craft and equip a stone pickaxe, build a furnace from cobblestone, mine iron ore, smelt it into iron ingots, then craft the iron pickaxe.
RULES (Minecraft facts, always true):
- 1 log -> 4 planks (craft in your 2x2 grid, no table). 4 planks -> crafting table. 2 planks -> 4 sticks.
- Wooden pickaxe = 3 planks + 2 sticks, at a crafting table.
- Stone/cobblestone drops ONLY while a pickaxe is EQUIPPED (bare hands get nothing).
- Stone pickaxe = 3 cobblestone + 2 sticks, at the table. Furnace = 8 cobblestone.
- Iron ore drops ONLY while a STONE (or better) pickaxe is EQUIPPED; a wooden pickaxe breaks it but drops NOTHING.
- Iron ore is found underground, mostly y=10..40 (you see your y in World).
- Smelting: with the furnace and fuel (planks or coal) in your inventory, say "smelt iron" -> 1 iron ingot per ore.
- Iron pickaxe = 3 iron ingots + 2 sticks, at the table.

Decision 102/120

Totals so far: coal_ore x2, crafting_table x3, diorite x1, dirt x3, grass_block x1, spruce_leaves x7, spruce_log x5, stone x14

World: y=50 (on the surface, light 15/15), facing north, 12127 blocks from where you started
Holding: stone_pickaxe

What you have already done (...95 earlier turns omitted; most recent last):
  96. "tunnel forward looking for iron" -> got nothing new
  97. "tunnel forward looking for iron" -> got nothing new
  98. "tunnel forward looking for iron" -> got nothing new
  99. "mine the ore in the cave wall" -> got stone x1, diorite x1, coal_ore x1
  100. "mine the ore in the cave wall" -> got stone x1
  101. "mine the ore in the cave wall" -> got nothing new
Inventory: coal x1, cobblestone x1, crafting_table x1, dirt x4, furnace x1, oak_planks x1, stone_pickaxe x1, wooden_pickaxe x1
Craftable now: nothing; missing: planks (need 1 log), sticks (need 1 planks)
Last: executed 'mine the ore in the cave wall' -> got nothing new
Reply with <think>your reasoning</think> then <action>one short instruction</action>.
\end{promptbox}

\textbf{Bidding Prompts.}

\begin{promptbox}[mblueL]{mblueD}{Logger Agent: Bid}{box:ec_bid_logger}
# Screenshot
<image>

TASK: Obtain an iron pickaxe. You start with nothing. Chop trees for logs, craft planks, a crafting table and sticks, craft and equip a wooden pickaxe, dig down and mine cobblestone, craft and equip a stone pickaxe, build a furnace from cobblestone, mine iron ore, smelt it into iron ingots, then craft the iron pickaxe.
RULES (Minecraft facts, always true):
- 1 log -> 4 planks (craft in your 2x2 grid, no table). 4 planks -> crafting table. 2 planks -> 4 sticks.
- Wooden pickaxe = 3 planks + 2 sticks, at a crafting table.
- Stone/cobblestone drops ONLY while a pickaxe is EQUIPPED (bare hands get nothing).
- Stone pickaxe = 3 cobblestone + 2 sticks, at the table. Furnace = 8 cobblestone.
- Iron ore drops ONLY while a STONE (or better) pickaxe is EQUIPPED; a wooden pickaxe breaks it but drops NOTHING.
- Iron ore is found underground, mostly y=10..40 (you see your y in World).
- Smelting: with the furnace and fuel (planks or coal) in your inventory, say "smelt iron" -> 1 iron ingot per ore.
- Iron pickaxe = 3 iron ingots + 2 sticks, at the table.

Decision 102/120

Totals so far: coal_ore x2, crafting_table x3, diorite x1, dirt x3, grass_block x1, spruce_leaves x7, spruce_log x5, stone x14

World: y=50 (on the surface, light 15/15), facing north, 12127 blocks from where you started
Holding: stone_pickaxe

What you have already done (...95 earlier turns omitted; most recent last):
  96. "tunnel forward looking for iron" -> got nothing new
  97. "tunnel forward looking for iron" -> got nothing new
  98. "tunnel forward looking for iron" -> got nothing new
  99. "mine the ore in the cave wall" -> got stone x1, diorite x1, coal_ore x1
  100. "mine the ore in the cave wall" -> got stone x1
  101. "mine the ore in the cave wall" -> got nothing new
Inventory: coal x1, cobblestone x1, crafting_table x1, dirt x4, furnace x1, oak_planks x1, stone_pickaxe x1, wooden_pickaxe x1
Craftable now: nothing; missing: planks (need 1 log), sticks (need 1 planks)
Last: executed 'mine the ore in the cave wall' -> got nothing new

# You are the LOGGER desk
You are one of 3 competing desks that share control of this single Minecraft player.
Exactly ONE desk acts each turn, decided by a sealed-bid auction that you are bidding in now.
If you win you must issue an instruction whose FIRST WORD is one of: chop | get | gather | break | find | scatter.
You cannot issue any other kind of instruction, so a turn you win that your verbs cannot
advance is a turn wasted for the whole team.

# Bidding (revenue share, five levels)
Your bid is the share of your income this turn that you promise to pass back to the desk
that acted before you (you are buying the state it produced). Your income is what the
environment pays your action plus the share the NEXT winner passes back to you, so value
flows backward along the whole chain. The HIGHEST bid wins the turn.
First read the Inventory and 'Craftable now' lines literally and decide whether YOUR verbs
can advance the iron-pickaxe chain from THIS exact state. Then bid:
  0    = pass. My verbs cannot help right now; another desk must act first.
  0.2  = my verbs barely help here.
  0.4  = my verbs are useful here.
  0.6  = my verbs are important here.
  0.8  = I am the bottleneck desk: the chain cannot move without my verbs right now.

Answer with ONE short sentence of reasoning that cites concrete inventory facts, then the
number, for example:
<think>one sentence about what is in the inventory and what my verbs can do with it</think>
<bid>0.4</bid>
\end{promptbox}

\begin{promptbox}[mblueL]{mblueD}{Miner Agent: Bid}{box:ec_bid_miner}
# Screenshot
<image>

TASK: Obtain an iron pickaxe. You start with nothing. Chop trees for logs, craft planks, a crafting table and sticks, craft and equip a wooden pickaxe, dig down and mine cobblestone, craft and equip a stone pickaxe, build a furnace from cobblestone, mine iron ore, smelt it into iron ingots, then craft the iron pickaxe.
RULES (Minecraft facts, always true):
- 1 log -> 4 planks (craft in your 2x2 grid, no table). 4 planks -> crafting table. 2 planks -> 4 sticks.
- Wooden pickaxe = 3 planks + 2 sticks, at a crafting table.
- Stone/cobblestone drops ONLY while a pickaxe is EQUIPPED (bare hands get nothing).
- Stone pickaxe = 3 cobblestone + 2 sticks, at the table. Furnace = 8 cobblestone.
- Iron ore drops ONLY while a STONE (or better) pickaxe is EQUIPPED; a wooden pickaxe breaks it but drops NOTHING.
- Iron ore is found underground, mostly y=10..40 (you see your y in World).
- Smelting: with the furnace and fuel (planks or coal) in your inventory, say "smelt iron" -> 1 iron ingot per ore.
- Iron pickaxe = 3 iron ingots + 2 sticks, at the table.

Decision 102/120

Totals so far: coal_ore x2, crafting_table x3, diorite x1, dirt x3, grass_block x1, spruce_leaves x7, spruce_log x5, stone x14

World: y=50 (on the surface, light 15/15), facing north, 12127 blocks from where you started
Holding: stone_pickaxe

What you have already done (...95 earlier turns omitted; most recent last):
  96. "tunnel forward looking for iron" -> got nothing new
  97. "tunnel forward looking for iron" -> got nothing new
  98. "tunnel forward looking for iron" -> got nothing new
  99. "mine the ore in the cave wall" -> got stone x1, diorite x1, coal_ore x1
  100. "mine the ore in the cave wall" -> got stone x1
  101. "mine the ore in the cave wall" -> got nothing new
Inventory: coal x1, cobblestone x1, crafting_table x1, dirt x4, furnace x1, oak_planks x1, stone_pickaxe x1, wooden_pickaxe x1
Craftable now: nothing; missing: planks (need 1 log), sticks (need 1 planks)
Last: executed 'mine the ore in the cave wall' -> got nothing new

# You are the MINER desk
You are one of 3 competing desks that share control of this single Minecraft player.
Exactly ONE desk acts each turn, decided by a sealed-bid auction that you are bidding in now.
If you win you must issue an instruction whose FIRST WORD is one of: dig | mine | explore | tunnel.
You cannot issue any other kind of instruction, so a turn you win that your verbs cannot
advance is a turn wasted for the whole team.

# Bidding (revenue share, five levels)
Your bid is the share of your income this turn that you promise to pass back to the desk
that acted before you (you are buying the state it produced). Your income is what the
environment pays your action plus the share the NEXT winner passes back to you, so value
flows backward along the whole chain. The HIGHEST bid wins the turn.
First read the Inventory and 'Craftable now' lines literally and decide whether YOUR verbs
can advance the iron-pickaxe chain from THIS exact state. Then bid:
  0    = pass. My verbs cannot help right now; another desk must act first.
  0.2  = my verbs barely help here.
  0.4  = my verbs are useful here.
  0.6  = my verbs are important here.
  0.8  = I am the bottleneck desk: the chain cannot move without my verbs right now.

Answer with ONE short sentence of reasoning that cites concrete inventory facts, then the
number, for example:
<think>one sentence about what is in the inventory and what my verbs can do with it</think>
<bid>0.4</bid>
\end{promptbox}

\begin{promptbox}[mblueL]{mblueD}{Smith Agent: Bid}{box:ec_bid_smith}
# Screenshot
<image>

TASK: Obtain an iron pickaxe. You start with nothing. Chop trees for logs, craft planks, a crafting table and sticks, craft and equip a wooden pickaxe, dig down and mine cobblestone, craft and equip a stone pickaxe, build a furnace from cobblestone, mine iron ore, smelt it into iron ingots, then craft the iron pickaxe.
RULES (Minecraft facts, always true):
- 1 log -> 4 planks (craft in your 2x2 grid, no table). 4 planks -> crafting table. 2 planks -> 4 sticks.
- Wooden pickaxe = 3 planks + 2 sticks, at a crafting table.
- Stone/cobblestone drops ONLY while a pickaxe is EQUIPPED (bare hands get nothing).
- Stone pickaxe = 3 cobblestone + 2 sticks, at the table. Furnace = 8 cobblestone.
- Iron ore drops ONLY while a STONE (or better) pickaxe is EQUIPPED; a wooden pickaxe breaks it but drops NOTHING.
- Iron ore is found underground, mostly y=10..40 (you see your y in World).
- Smelting: with the furnace and fuel (planks or coal) in your inventory, say "smelt iron" -> 1 iron ingot per ore.
- Iron pickaxe = 3 iron ingots + 2 sticks, at the table.

Decision 102/120

Totals so far: coal_ore x2, crafting_table x3, diorite x1, dirt x3, grass_block x1, spruce_leaves x7, spruce_log x5, stone x14

World: y=50 (on the surface, light 15/15), facing north, 12127 blocks from where you started
Holding: stone_pickaxe

What you have already done (...95 earlier turns omitted; most recent last):
  96. "tunnel forward looking for iron" -> got nothing new
  97. "tunnel forward looking for iron" -> got nothing new
  98. "tunnel forward looking for iron" -> got nothing new
  99. "mine the ore in the cave wall" -> got stone x1, diorite x1, coal_ore x1
  100. "mine the ore in the cave wall" -> got stone x1
  101. "mine the ore in the cave wall" -> got nothing new
Inventory: coal x1, cobblestone x1, crafting_table x1, dirt x4, furnace x1, oak_planks x1, stone_pickaxe x1, wooden_pickaxe x1
Craftable now: nothing; missing: planks (need 1 log), sticks (need 1 planks)
Last: executed 'mine the ore in the cave wall' -> got nothing new

# You are the SMITH desk
You are one of 3 competing desks that share control of this single Minecraft player.
Exactly ONE desk acts each turn, decided by a sealed-bid auction that you are bidding in now.
If you win you must issue an instruction whose FIRST WORD is one of: craft | equip | smelt.
You cannot issue any other kind of instruction, so a turn you win that your verbs cannot
advance is a turn wasted for the whole team.

# Bidding (revenue share, five levels)
Your bid is the share of your income this turn that you promise to pass back to the desk
that acted before you (you are buying the state it produced). Your income is what the
environment pays your action plus the share the NEXT winner passes back to you, so value
flows backward along the whole chain. The HIGHEST bid wins the turn.
First read the Inventory and 'Craftable now' lines literally and decide whether YOUR verbs
can advance the iron-pickaxe chain from THIS exact state. Then bid:
  0    = pass. My verbs cannot help right now; another desk must act first.
  0.2  = my verbs barely help here.
  0.4  = my verbs are useful here.
  0.6  = my verbs are important here.
  0.8  = I am the bottleneck desk: the chain cannot move without my verbs right now.

Answer with ONE short sentence of reasoning that cites concrete inventory facts, then the
number, for example:
<think>one sentence about what is in the inventory and what my verbs can do with it</think>
<bid>0.4</bid>
\end{promptbox}

\textbf{Action Prompts.}

\begin{promptbox}[mblueL]{mblueD}{Logger Agent: Act}{box:ec_act_logger}
# Screenshot
<image>

TASK: Obtain an iron pickaxe. You start with nothing. Chop trees for logs, craft planks, a crafting table and sticks, craft and equip a wooden pickaxe, dig down and mine cobblestone, craft and equip a stone pickaxe, build a furnace from cobblestone, mine iron ore, smelt it into iron ingots, then craft the iron pickaxe.
RULES (Minecraft facts, always true):
- 1 log -> 4 planks (craft in your 2x2 grid, no table). 4 planks -> crafting table. 2 planks -> 4 sticks.
- Wooden pickaxe = 3 planks + 2 sticks, at a crafting table.
- Stone/cobblestone drops ONLY while a pickaxe is EQUIPPED (bare hands get nothing).
- Stone pickaxe = 3 cobblestone + 2 sticks, at the table. Furnace = 8 cobblestone.
- Iron ore drops ONLY while a STONE (or better) pickaxe is EQUIPPED; a wooden pickaxe breaks it but drops NOTHING.
- Iron ore is found underground, mostly y=10..40 (you see your y in World).
- Smelting: with the furnace and fuel (planks or coal) in your inventory, say "smelt iron" -> 1 iron ingot per ore.
- Iron pickaxe = 3 iron ingots + 2 sticks, at the table.

Decision 102/120

Totals so far: coal_ore x2, crafting_table x3, diorite x1, dirt x3, grass_block x1, spruce_leaves x7, spruce_log x5, stone x14

World: y=50 (on the surface, light 15/15), facing north, 12127 blocks from where you started
Holding: stone_pickaxe

What you have already done (...95 earlier turns omitted; most recent last):
  96. "tunnel forward looking for iron" -> got nothing new
  97. "tunnel forward looking for iron" -> got nothing new
  98. "tunnel forward looking for iron" -> got nothing new
  99. "mine the ore in the cave wall" -> got stone x1, diorite x1, coal_ore x1
  100. "mine the ore in the cave wall" -> got stone x1
  101. "mine the ore in the cave wall" -> got nothing new
Inventory: coal x1, cobblestone x1, crafting_table x1, dirt x4, furnace x1, oak_planks x1, stone_pickaxe x1, wooden_pickaxe x1
Craftable now: nothing; missing: planks (need 1 log), sticks (need 1 planks)
Last: executed 'mine the ore in the cave wall' -> got nothing new
Reply with <think>your reasoning</think> then <action>one short instruction</action>.

# You are the LOGGER desk and you WON this turn's auction
You issue the single instruction for this turn. A low-level controller (STEVE-1) executes it
for about two seconds of game time; the crafting verbs are executed exactly by the server.

HARD CONSTRAINT, enforced by a parser and not negotiable: the FIRST WORD of your instruction
must be one of: chop | get | gather | break | find | scatter
Anything else is REJECTED before it reaches the game: the turn is wasted, nothing happens,
and you still pay your promised share. You are the surface-gathering desk: wood and anything reachable by walking up to it and breaking it. 'scatter' teleports the player a short distance -- use it only when the spawn has nothing usable in sight.

Keep it short, concrete and visual. Example: <action>chop the tree in front of you</action>
Reply with your reasoning, then the instruction wrapped EXACTLY like this:
<action>your short instruction here</action>
\end{promptbox}

\begin{promptbox}[mblueL]{mblueD}{Miner Agent: Act}{box:ec_act_miner}
# Screenshot
<image>

TASK: Obtain an iron pickaxe. You start with nothing. Chop trees for logs, craft planks, a crafting table and sticks, craft and equip a wooden pickaxe, dig down and mine cobblestone, craft and equip a stone pickaxe, build a furnace from cobblestone, mine iron ore, smelt it into iron ingots, then craft the iron pickaxe.
RULES (Minecraft facts, always true):
- 1 log -> 4 planks (craft in your 2x2 grid, no table). 4 planks -> crafting table. 2 planks -> 4 sticks.
- Wooden pickaxe = 3 planks + 2 sticks, at a crafting table.
- Stone/cobblestone drops ONLY while a pickaxe is EQUIPPED (bare hands get nothing).
- Stone pickaxe = 3 cobblestone + 2 sticks, at the table. Furnace = 8 cobblestone.
- Iron ore drops ONLY while a STONE (or better) pickaxe is EQUIPPED; a wooden pickaxe breaks it but drops NOTHING.
- Iron ore is found underground, mostly y=10..40 (you see your y in World).
- Smelting: with the furnace and fuel (planks or coal) in your inventory, say "smelt iron" -> 1 iron ingot per ore.
- Iron pickaxe = 3 iron ingots + 2 sticks, at the table.

Decision 102/120

Totals so far: coal_ore x2, crafting_table x3, diorite x1, dirt x3, grass_block x1, spruce_leaves x7, spruce_log x5, stone x14

World: y=50 (on the surface, light 15/15), facing north, 12127 blocks from where you started
Holding: stone_pickaxe

What you have already done (...95 earlier turns omitted; most recent last):
  96. "tunnel forward looking for iron" -> got nothing new
  97. "tunnel forward looking for iron" -> got nothing new
  98. "tunnel forward looking for iron" -> got nothing new
  99. "mine the ore in the cave wall" -> got stone x1, diorite x1, coal_ore x1
  100. "mine the ore in the cave wall" -> got stone x1
  101. "mine the ore in the cave wall" -> got nothing new
Inventory: coal x1, cobblestone x1, crafting_table x1, dirt x4, furnace x1, oak_planks x1, stone_pickaxe x1, wooden_pickaxe x1
Craftable now: nothing; missing: planks (need 1 log), sticks (need 1 planks)
Last: executed 'mine the ore in the cave wall' -> got nothing new
Reply with <think>your reasoning</think> then <action>one short instruction</action>.

# You are the MINER desk and you WON this turn's auction
You issue the single instruction for this turn. A low-level controller (STEVE-1) executes it
for about two seconds of game time; the crafting verbs are executed exactly by the server.

HARD CONSTRAINT, enforced by a parser and not negotiable: the FIRST WORD of your instruction
must be one of: dig | mine | explore | tunnel
Anything else is REJECTED before it reaches the game: the turn is wasted, nothing happens,
and you still pay your promised share. You are the underground desk: getting below the surface and mining stone and ores.

Keep it short, concrete and visual. Example: <action>dig down to find stone</action>
Reply with your reasoning, then the instruction wrapped EXACTLY like this:
<action>your short instruction here</action>
\end{promptbox}

\begin{promptbox}[mblueL]{mblueD}{Smith Agent: Act}{box:ec_act_smith}
# Screenshot
<image>

TASK: Obtain an iron pickaxe. You start with nothing. Chop trees for logs, craft planks, a crafting table and sticks, craft and equip a wooden pickaxe, dig down and mine cobblestone, craft and equip a stone pickaxe, build a furnace from cobblestone, mine iron ore, smelt it into iron ingots, then craft the iron pickaxe.
RULES (Minecraft facts, always true):
- 1 log -> 4 planks (craft in your 2x2 grid, no table). 4 planks -> crafting table. 2 planks -> 4 sticks.
- Wooden pickaxe = 3 planks + 2 sticks, at a crafting table.
- Stone/cobblestone drops ONLY while a pickaxe is EQUIPPED (bare hands get nothing).
- Stone pickaxe = 3 cobblestone + 2 sticks, at the table. Furnace = 8 cobblestone.
- Iron ore drops ONLY while a STONE (or better) pickaxe is EQUIPPED; a wooden pickaxe breaks it but drops NOTHING.
- Iron ore is found underground, mostly y=10..40 (you see your y in World).
- Smelting: with the furnace and fuel (planks or coal) in your inventory, say "smelt iron" -> 1 iron ingot per ore.
- Iron pickaxe = 3 iron ingots + 2 sticks, at the table.

Decision 102/120

Totals so far: coal_ore x2, crafting_table x3, diorite x1, dirt x3, grass_block x1, spruce_leaves x7, spruce_log x5, stone x14

World: y=50 (on the surface, light 15/15), facing north, 12127 blocks from where you started
Holding: stone_pickaxe

What you have already done (...95 earlier turns omitted; most recent last):
  96. "tunnel forward looking for iron" -> got nothing new
  97. "tunnel forward looking for iron" -> got nothing new
  98. "tunnel forward looking for iron" -> got nothing new
  99. "mine the ore in the cave wall" -> got stone x1, diorite x1, coal_ore x1
  100. "mine the ore in the cave wall" -> got stone x1
  101. "mine the ore in the cave wall" -> got nothing new
Inventory: coal x1, cobblestone x1, crafting_table x1, dirt x4, furnace x1, oak_planks x1, stone_pickaxe x1, wooden_pickaxe x1
Craftable now: nothing; missing: planks (need 1 log), sticks (need 1 planks)
Last: executed 'mine the ore in the cave wall' -> got nothing new
Reply with <think>your reasoning</think> then <action>one short instruction</action>.

# You are the SMITH desk and you WON this turn's auction
You issue the single instruction for this turn. A low-level controller (STEVE-1) executes it
for about two seconds of game time; the crafting verbs are executed exactly by the server.

HARD CONSTRAINT, enforced by a parser and not negotiable: the FIRST WORD of your instruction
must be one of: craft | equip | smelt
Anything else is REJECTED before it reaches the game: the turn is wasted, nothing happens,
and you still pay your promised share. You are the inventory desk: recipes and equipping. These verbs are executed exactly by the server, so a correctly-timed craft never fails for motor reasons -- but a craft with the ingredients missing fails outright and wastes the turn.

Keep it short, concrete and visual. Example: <action>craft a stone pickaxe</action>
Reply with your reasoning, then the instruction wrapped EXACTLY like this:
<action>your short instruction here</action>
\end{promptbox}

\subsubsection{SFT Cold Start}
To teach the models the input and output formats and establish basic task capabilities, we perform an SFT cold start in both settings. Specifically, we use Claude Sonnet 5 to generate demonstrations of crafting-task execution and record all model responses. We collect 4,791 examples from actual Minecraft interactions, including 1,677 for \textsc{Logger}, 2,380 for \textsc{Miner}, and 734 for \textsc{Smith}. We fine-tune the monolithic agent on the full dataset and each society agent on its corresponding subset, matching the total training compute between the two settings during cold start. SFT for each model is performed on a single B200 GPU.

We fine-tune all model parameters in bf16 using AdamW with a cosine learning rate schedule. Detailed settings are listed in Table~\ref{tab:hyperparams_ec}.

\subsubsection{Post-training}
Our post-training dataset consists of 1,614 random seeds, each generating a different Minecraft world. We host the Minecraft environments and the STEVE-1 model on four RTX PRO 6000 GPUs, allowing 32 Minecraft worlds to run in parallel. Reinforcement learning is performed on four B200 GPUs. Detailed settings are listed in Table~\ref{tab:hyperparams_ec}.

\begin{table*}[h]
\centering
\caption{Hyperparameters for the Embodied Crafting experiment.}
\label{tab:hyperparams_ec}
\begin{tabularx}{\textwidth}{@{}lX@{}}
\toprule
\textbf{Category} & \textbf{Hyperparameter $=$ Value} \\
\midrule
\textbf{SFT cold start} &
epochs $=$ 2; per\_device\_batch $=$ 4; grad\_accum $=$ 4; learning\_rate $=$ $1\!\times\!10^{-5}$; warmup\_ratio $=$ 0.03; weight\_decay $=$ 0; max\_grad\_norm $=$ 1.0. \\
\midrule
\textbf{Post-training} &
learning\_rate $=$ $1\!\times\!10^{-6}$; train\_batch\_size $=$ 32;  grad\_clip $=$ 1.0; entropy\_coeff $=$ 0. \\
\midrule
\textbf{Generation} &
temperature $=$ 1.0; top\_p $=$ 1.0; \\
& max\_prompt\_length $=$ 3{,}072; max\_response\_length $=$ 320 (action), 96 (bid).\\
\midrule
\textbf{Environment} &
decisions\_per\_episode $=$ 120; frames\_per\_instruction $=$ 40. \\
\bottomrule
\end{tabularx}
\end{table*}

\subsection{Robot Navigation}
\label{app:task_rn}

\subsubsection{Task and Simulator}
The task is vision-and-language navigation in continuous environments (VLN-CE~\citep{vlnce}): given a natural-language route instruction from R2R~\citep{r2r}, the agent must move through a photorealistic Matterport3D house~\citep{mp3d} in the Habitat simulator~\citep{habitat} and stop within 3\,m of the goal. We use the NaVILA evaluation setup~\citep{navila}, in which the simulator's atomic actions are a 0.25\,m forward step and a 15$^\circ$ turn. 

\subsubsection{Reward}
We use the VLN-CE waypoint reward. Let $d_k$ be the geodesic distance from the agent to the goal after the $k$-th atomic step of the episode, $d_0$ the distance at the start, and let decision $t$ expand into the atomic steps $k\in\mathcal{K}_t$. The reward at decision $t$ and the episode return are
\begin{equation}
r_t=\sum_{k\in\mathcal{K}_t}\Big[\,\lambda_{\mathrm{d}}\,(d_{k-1}-d_k)+\lambda_{\mathrm{s}}\,\mathbf{1}[\text{success at }k]\,\Big],\qquad
R=\sum_{t} r_t=\lambda_{\mathrm{d}}\,(d_0-d_{K})+\lambda_{\mathrm{s}}\,\mathbf{1}[\text{success}],
\label{eq:rn_reward}
\end{equation}
with $\lambda_{\mathrm{d}}=1.0$ and $\lambda_{\mathrm{s}}=2.5$, where success means the agent issued \emph{stop} within 3\,m of the goal and $K$ is the last atomic step.

\subsubsection{Episode, Observation and Decision}
An episode contains at most 64 decisions. At each decision the policy receives the instruction and an eight-frame window of $512\times512$ egocentric RGB frames: the current frame plus seven frames sampled uniformly from the history, padded with black frames at the start of the episode. The policy answers with reasoning followed by one mid-level action, \texttt{<think>...</think><act>...</act>}, where the action is \emph{turn left} or \emph{turn right} by 15, 30 or 45 degrees, \emph{move forward} by 25, 50 or 75\,cm, or \emph{stop}. 

\subsubsection{Agent Design}
The population consists of three agents, \textsc{Turner}, \textsc{Mover} and \textsc{Stopper}, each a separate full fine-tune of NaVILA-8B~\citep{navila}. They see the same frames and instruction and partition the action space: \textsc{Turner} owns the six turning actions, \textsc{Mover} the three forward moves, and \textsc{Stopper} the \emph{stop}.
The monolithic agent is the same base model fine-tuned on the full action space.
Below, we present the prompt for the monolithic agent and the bidding and action prompts for each agent in the society.

\textbf{Single Agent.}

\begin{promptbox}[mblueL]{mblueD}{Single Agent}{box:rn_single}
Imagine you are a robot programmed for navigation tasks. You have been given a video of historical observations <image>
<image>
<image>
<image>
<image>
<image>
<image>
, and current observation <image>
. Your assigned task is: "Walk straight along a corridor containing two glass walls with frosted glass in the middle. Take the first door on the left and wait inside the door." Analyze this series of images to decide your next action, which could be turning left or right by a specific degree, moving forward a certain distance, or stop if the task is completed.
\end{promptbox}

\textbf{Bidding Prompts.}

\begin{promptbox}[mblueL]{mblueD}{Turner Agent: Bid}{box:rn_bid_turner}
Imagine you are a robot programmed for navigation tasks. You have been given a video of historical observations <image>
<image>
<image>
<image>
<image>
<image>
<image>
, and current observation <image>
. Your assigned task is: "Walk straight along a corridor containing two glass walls with frosted glass in the middle. Take the first door on the left and wait inside the door." You are the TURNER desk, one of 3 specialists (Turner/Mover/Stopper) that share control of this robot. Exactly ONE desk acts each step, decided by a sealed-bid auction you are bidding in now. If you win you may ONLY emit: turn left or turn right by 15, 30, or 45 degrees. You cannot emit anything else, so bid only for steps your action can advance.
# Bidding (revenue share, five levels)
Your bid is the share of your income this step you pass back to the desk that acted before you. The HIGHEST bid wins. Read the frames and the instruction, and bid:
  0    = pass. My action cannot help now; another desk must act first.
  0.2  = my action barely helps here.
  0.4  = my action is useful here.
  0.6  = my action is important here.
  0.8  = I am the bottleneck: the robot cannot progress without my action right now.
 Answer with ONE short sentence of reasoning about the scene, then the number:
<think>one sentence</think>
<bid>0.4</bid>
\end{promptbox}

\begin{promptbox}[mblueL]{mblueD}{Mover Agent: Bid}{box:rn_bid_mover}
Imagine you are a robot programmed for navigation tasks. You have been given a video of historical observations <image>
<image>
<image>
<image>
<image>
<image>
<image>
, and current observation <image>
. Your assigned task is: "Walk straight along a corridor containing two glass walls with frosted glass in the middle. Take the first door on the left and wait inside the door." You are the MOVER desk, one of 3 specialists (Turner/Mover/Stopper) that share control of this robot. Exactly ONE desk acts each step, decided by a sealed-bid auction you are bidding in now. If you win you may ONLY emit: move forward by 25, 50, or 75 cm. You cannot emit anything else, so bid only for steps your action can advance.
# Bidding (revenue share, five levels)
Your bid is the share of your income this step you pass back to the desk that acted before you. The HIGHEST bid wins. Read the frames and the instruction, and bid:
  0    = pass. My action cannot help now; another desk must act first.
  0.2  = my action barely helps here.
  0.4  = my action is useful here.
  0.6  = my action is important here.
  0.8  = I am the bottleneck: the robot cannot progress without my action right now.
 Answer with ONE short sentence of reasoning about the scene, then the number:
<think>one sentence</think>
<bid>0.4</bid>
\end{promptbox}

\begin{promptbox}[mblueL]{mblueD}{Stopper Agent: Bid}{box:rn_bid_stopper}
Imagine you are a robot programmed for navigation tasks. You have been given a video of historical observations <image>
<image>
<image>
<image>
<image>
<image>
<image>
, and current observation <image>
. Your assigned task is: "Walk straight along a corridor containing two glass walls with frosted glass in the middle. Take the first door on the left and wait inside the door." You are the STOPPER desk, one of 3 specialists (Turner/Mover/Stopper) that share control of this robot. Exactly ONE desk acts each step, decided by a sealed-bid auction you are bidding in now. If you win you may ONLY emit: stop. You cannot emit anything else, so bid only for steps your action can advance.
# Bidding (revenue share, five levels)
Your bid is the share of your income this step you pass back to the desk that acted before you. The HIGHEST bid wins. Read the frames and the instruction, and bid:
  0    = pass. My action cannot help now; another desk must act first.
  0.2  = my action barely helps here.
  0.4  = my action is useful here.
  0.6  = my action is important here.
  0.8  = I am the bottleneck: the robot cannot progress without my action right now.
 Answer with ONE short sentence of reasoning about the scene, then the number:
<think>one sentence</think>
<bid>0.4</bid>
\end{promptbox}

\textbf{Action Prompts.}

\begin{promptbox}[mblueL]{mblueD}{Turner Agent: Act}{box:rn_act_turner}
Imagine you are a robot programmed for navigation tasks. You have been given a video of historical observations <image>
<image>
<image>
<image>
<image>
<image>
<image>
, and current observation <image>
. Your assigned task is: "Walk straight along a corridor containing two glass walls with frosted glass in the middle. Take the first door on the left and wait inside the door." You are the TURNER desk and you WON this step's auction. You issue the single navigation action for this step. HARD CONSTRAINT, enforced by a parser: your action must be turn left or turn right by 15, 30, or 45 degrees. Anything else is REJECTED, the step is wasted, and you still pay. You orient the robot. Use a turn to face the direction the instruction wants before the Mover advances. You cannot move forward or stop.
Analyze the images to decide your action.
\end{promptbox}

\begin{promptbox}[mblueL]{mblueD}{Mover Agent: Act}{box:rn_act_mover}
Imagine you are a robot programmed for navigation tasks. You have been given a video of historical observations <image>
<image>
<image>
<image>
<image>
<image>
<image>
, and current observation <image>
. Your assigned task is: "Walk straight along a corridor containing two glass walls with frosted glass in the middle. Take the first door on the left and wait inside the door." You are the MOVER desk and you WON this step's auction. You issue the single navigation action for this step. HARD CONSTRAINT, enforced by a parser: your action must be move forward by 25, 50, or 75 cm. Anything else is REJECTED, the step is wasted, and you still pay. You make forward progress once the robot faces the right way. You cannot turn or stop.
Analyze the images to decide your action.
\end{promptbox}

\begin{promptbox}[mblueL]{mblueD}{Stopper Agent: Act}{box:rn_act_stopper}
Imagine you are a robot programmed for navigation tasks. You have been given a video of historical observations <image>
<image>
<image>
<image>
<image>
<image>
<image>
, and current observation <image>
. Your assigned task is: "Walk straight along a corridor containing two glass walls with frosted glass in the middle. Take the first door on the left and wait inside the door." You are the STOPPER desk and you WON this step's auction. You issue the single navigation action for this step. HARD CONSTRAINT, enforced by a parser: your action must be stop. Anything else is REJECTED, the step is wasted, and you still pay. You end the episode by stopping. Win ONLY when the robot has reached the goal the instruction describes -- stopping early or late both fail. You cannot turn or move.
Analyze the images to decide your action.
\end{promptbox}

\subsubsection{SFT Cold Start}
To teach the models the input and output formats and establish basic navigation capability, we perform an SFT cold start in both settings. Starting from the multimodal-alignment checkpoint of NaVILA, we render oracle trajectories on R2R-CE training episodes with Habitat's shortest-path follower. We keep successful episodes containing 1{,}931 decisions (800 turns, 800 moves and 331 stops), and use Claude Sonnet 5 to annotate data. 
We fine-tune the monolithic agent on all data and each society agent on its own data. SFT for each model is performed on a single B200 GPU.

We fine-tune all model parameters in bf16 using AdamW with a cosine learning rate schedule. Detailed settings are listed in Table~\ref{tab:hyperparams_rn}.

\subsubsection{Post-training}
Our post-training dataset consists of 10{,}744 R2R-CE training episodes different from SFT. We host the Habitat environments on 2 RTX PRO 6000 GPUs. Reinforcement learning is performed on 4 B200 GPUs. Detailed settings are listed in Table~\ref{tab:hyperparams_rn}.

\begin{table*}[h]
\centering
\caption{Hyperparameters for the Robot Navigation experiment.}
\label{tab:hyperparams_rn}
\begin{tabularx}{\textwidth}{@{}lX@{}}
\toprule
\textbf{Category} & \textbf{Hyperparameter $=$ Value} \\
\midrule
\textbf{SFT cold start} &
epochs $=$ 2; per\_device\_batch $=$ 2; grad\_accum $=$ 8; learning\_rate $=$ $2\!\times\!10^{-5}$; warmup\_ratio $=$ 0.03; weight\_decay $=$ 0; max\_grad\_norm $=$ 1.0. \\
\midrule
\textbf{Post-training} &
learning\_rate $=$ $1\!\times\!10^{-6}$; train\_batch\_size $=$ 16;  grad\_clip $=$ 1.0; entropy\_coeff $=$ 0. \\
\midrule
\textbf{Generation} &
temperature $=$ 1.0; top\_p $=$ 1.0; num\_frames $=$ 8; \\
& max\_prompt\_length $=$ 4{,}096; max\_response\_length $=$ 96. \\
\midrule
\textbf{Environment} &
decisions\_per\_episode $=$ 64;  frame\_size $=$ $512\times512$. \\
\bottomrule
\end{tabularx}
\end{table*}

\subsection{Deep Research}
\label{app:task_dr}

\subsubsection{Task and Environment}
The task is open-domain deep research: given a question, the agent must browse a local web corpus with three tools and return a short final answer. We use the OpenResearcher benchmark~\citep{openresearcher}: the questions are the OpenResearcher-Dataset, and the corpus is the OpenResearcher-Corpus, served by a BM25 search service.  Each question comes with one reference answer.

The browser exposes OpenResearcher's own three tools, called from the model output as JSON: \texttt{browser.search} (a query, returning the top-10 hits as title, URL and summary), \texttt{browser.open} (a link id from the current page, a URL, or a page cursor, with \texttt{loc}/\texttt{num\_lines} to scroll) and \texttt{browser.find} (an exact-text pattern inside the currently open page).

\subsubsection{Reward}
The environment gives a single terminal reward. It is graded by an LLM judge (\texttt{qwen3.5-flash}) with OpenResearcher's HLE grading template, which extracts the final answer from the response and returns \emph{yes}/\emph{no} against the reference answer.  Let $T$ be the answering turn; the reward at decision $t$ and the episode return are
\begin{equation}
r_t=\mathbf{1}[t=T]\cdot\mathbf{1}[\text{judge says the answer is correct}],\qquad
R=\sum_{t} r_t\in\{0,1\}.
\label{eq:dr_reward}
\end{equation}

\subsubsection{Episode, Observation and Decision}
An episode contains at most 64 decisions. The observation is the full conversation so far: the system prompt (OpenResearcher's browsing instructions, the date, and the three tool schemas), the question, and the alternating assistant turns and tool responses. The prompt budget is 16{,}384 tokens. The policy answers with reasoning followed by either one tool call or the final answer, \texttt{<think>...</think><tool\_call>\{...\}</tool\_call>} or \texttt{<think>...</think><final\_answer>\{...\}</final\_answer>}. Generation stops at \texttt{<tool\_response>}, and the tool call is executed by the browser against the search service.

\subsubsection{Agent Design}
The population consists of three agents, \textsc{Searcher}, \textsc{Opener} and \textsc{Finder}, each a separate full fine-tune of gemma-4-E2B-it~\citep{gemma4}. They see the same conversation and partition the tool space: \textsc{Searcher} owns \texttt{browser.search}, \textsc{Opener} owns \texttt{browser.open}, and \textsc{Finder} owns \texttt{browser.find}; any agent may end the episode by giving the final answer. 
The monolithic agent is the same base model fine-tuned on turns spanning all three tools.

Below, we present the example prompt for the monolithic agent and the example bidding and action prompts for each agent in the society. 

\textbf{Single Agent.}

\begin{promptbox}[mblueL]{mblueD}{Single Agent}{box:dr_single}
[system]
You are a helpful assistant and harmless assistant.

You will be able to use a set of browsering tools to answer user queries.

Tool for browsing.
The `cursor` appears in brackets before each browsing display: `[{cursor}]`.
Cite information from the tool using the following format:
`[{cursor}\dagger L{line_start}(-L{line_end})?]`, for example: `[6\dagger L9-L11]` or `[8\dagger L3]`.
Do not quote more than 10 words directly from the tool output.
sources=web

Today's date: 2026-09-15

# Tools

You may call one or more functions to assist with the user query.

You are provided with function signatures within <tools></tools> XML tags:
<tools>
{"type": "function", "function": {"name": "browser.search", "description": "Searches for information related to a query and displays top N results. Returns a list of search results with titles, URLs, and summaries.", "parameters": {... "query": "The search query string", "topn": ...}}}
{"type": "function", "function": {"name": "browser.open", "description": "Opens a link from the current page or a fully qualified URL. Can scroll to a specific location and display a specific number of lines. Valid link ids are displayed with the formatting: [{id}\dagger .*].", "parameters": {... "id", "cursor", "loc", "num_lines" ...}}}
{"type": "function", "function": {"name": "browser.find", "description": "Finds exact matches of a pattern in the current page or a specified page by cursor.", "parameters": {... "pattern": "The exact text pattern to search for", "cursor" ...}}}
</tools>

For each function call, return a json object with function name and arguments within <tool_call></tool_call> XML tags:
<tool_call>
{"name": <function-name>, "arguments": <args-json-object>}
</tool_call>

[user]
Find the role John Wayne played in the 1969 film True Grit

[assistant]
<think>I should search for the film's cast.</think>
<tool_call>
{"name": "browser.search", "arguments": {"query": "John Wayne True Grit 1969 role"}}
</tool_call>

[user]
<tool_response>
[0] John Wayne True Grit 1969 role (web-search://ts=1757955200)
**viewing lines [0 - 22] of 22**

L0:
L1: URL: web-search://ts=1757955200
L2: # Search Results
L3:
L4:   * [0\dagger True Grit (1969 film)\dagger en.wikipedia.org] True Grit is a 1969 American Western film ... starring John Wayne as U.S. Marshal Rooster Cogburn ...
L5:   * [1\dagger Rooster Cogburn (character)\dagger en.wikipedia.org] Reuben J. "Rooster" Cogburn is a fictional character ...
  ... (8 more hits)
</tool_response>

[assistant]
<think>Hit 0 names the role; I will open it to confirm.</think>
<tool_call>
{"name": "browser.open", "arguments": {"id": 0}}
</tool_call>

[user]
<tool_response>
[1] True Grit (1969 film) (en.wikipedia.org/wiki/True_Grit_(1969_film))
**viewing lines [0 - 61] of 214**

L0:
L1: URL: https://en.wikipedia.org/wiki/True_Grit_(1969_film)
L2: True Grit is a 1969 American Western film directed by Henry Hathaway ...
L3: It stars John Wayne as U.S. Marshal Rooster Cogburn ...
  ...
</tool_response>
\end{promptbox}

\textbf{Bidding Prompts.} The system prompt is identical to the single agent's; the agent-specific paragraph is appended. We show the parts that differ.

\begin{promptbox}[mblueL]{mblueD}{Searcher Agent: Bid}{box:dr_bid_searcher}
[system, appended after the browsing instructions and before the tool schemas]
You are the SEARCHER desk, one of 3 specialists (Searcher/Opener/Finder) that share this browser and this conversation. Exactly one desk acts each turn, chosen by an auction. You may ONLY call browser.search, or give the final answer with no tool call. Any other tool call is rejected and the turn is wasted. You decide what to look for: craft queries that decompose the question into findable facts. You cannot open or read pages.

[user, the question and the conversation so far as in the single-agent prompt]
...

# You are the SEARCHER desk
You are one of 3 competing desks (Searcher/Opener/Finder) that share control of this research session. Exactly ONE desk acts each turn, decided by a sealed-bid auction that you are bidding in now. If you win you may only call browser.search, or give the final answer. A turn you win that your tools cannot advance is a turn wasted for the whole team.

# Bidding (revenue share, five levels)
Your bid is the share of your income this turn that you promise to pass back to the desk that acted before you (you are buying the state it produced). Your income is what the final answer earns plus the share the NEXT winner passes back to you, so value flows backward along the whole chain. The HIGHEST bid wins the turn.
Read the conversation so far literally -- what has been searched, which pages were opened, what evidence is already on screen -- and decide whether YOUR tools can advance the answer from THIS exact state. Then bid:
  0    = pass. My tools cannot help right now; another desk must act first.
  0.2  = my tools barely help here.
  0.4  = my tools are useful here.
  0.6  = my tools are important here.
  0.8  = I am the bottleneck desk: the answer cannot move without my tools right now.

Answer with ONE short sentence of reasoning that cites what is already in the conversation, then the number, for example:
<think>one sentence about the evidence so far and what my tools can do next</think>
<bid>0.4</bid>
\end{promptbox}

\begin{promptbox}[mblueL]{mblueD}{Opener Agent: Bid}{box:dr_bid_opener}
[system, appended after the browsing instructions and before the tool schemas]
You are the OPENER desk, one of 3 specialists (Searcher/Opener/Finder) that share this browser and this conversation. Exactly one desk acts each turn, chosen by an auction. You may ONLY call browser.open, or give the final answer with no tool call. Any other tool call is rejected and the turn is wasted. You open pages: follow link ids from the search results or open URLs directly, and scroll with loc / num_lines to read further. You cannot run searches or find text in a page.

[user, the question and the conversation so far as in the single-agent prompt]
...

# You are the OPENER desk
You are one of 3 competing desks (Searcher/Opener/Finder) that share control of this research session. Exactly ONE desk acts each turn, decided by a sealed-bid auction that you are bidding in now. If you win you may only call browser.open, or give the final answer. A turn you win that your tools cannot advance is a turn wasted for the whole team.

# Bidding (revenue share, five levels)
Your bid is the share of your income this turn that you promise to pass back to the desk that acted before you (you are buying the state it produced). Your income is what the final answer earns plus the share the NEXT winner passes back to you, so value flows backward along the whole chain. The HIGHEST bid wins the turn.
Read the conversation so far literally -- what has been searched, which pages were opened, what evidence is already on screen -- and decide whether YOUR tools can advance the answer from THIS exact state. Then bid:
  0    = pass. My tools cannot help right now; another desk must act first.
  0.2  = my tools barely help here.
  0.4  = my tools are useful here.
  0.6  = my tools are important here.
  0.8  = I am the bottleneck desk: the answer cannot move without my tools right now.

Answer with ONE short sentence of reasoning that cites what is already in the conversation, then the number, for example:
<think>one sentence about the evidence so far and what my tools can do next</think>
<bid>0.4</bid>
\end{promptbox}

\begin{promptbox}[mblueL]{mblueD}{Finder Agent: Bid}{box:dr_bid_finder}
[system, appended after the browsing instructions and before the tool schemas]
You are the FINDER desk, one of 3 specialists (Searcher/Opener/Finder) that share this browser and this conversation. Exactly one desk acts each turn, chosen by an auction. You may ONLY call browser.find, or give the final answer with no tool call. Any other tool call is rejected and the turn is wasted. You search inside the page that is currently open: find a pattern to confirm or rule out a candidate fact. You cannot run web searches or open pages.

[user, the question and the conversation so far as in the single-agent prompt]
...

# You are the FINDER desk
You are one of 3 competing desks (Searcher/Opener/Finder) that share control of this research session. Exactly ONE desk acts each turn, decided by a sealed-bid auction that you are bidding in now. If you win you may only call browser.find, or give the final answer. A turn you win that your tools cannot advance is a turn wasted for the whole team.

# Bidding (revenue share, five levels)
Your bid is the share of your income this turn that you promise to pass back to the desk that acted before you (you are buying the state it produced). Your income is what the final answer earns plus the share the NEXT winner passes back to you, so value flows backward along the whole chain. The HIGHEST bid wins the turn.
Read the conversation so far literally -- what has been searched, which pages were opened, what evidence is already on screen -- and decide whether YOUR tools can advance the answer from THIS exact state. Then bid:
  0    = pass. My tools cannot help right now; another desk must act first.
  0.2  = my tools barely help here.
  0.4  = my tools are useful here.
  0.6  = my tools are important here.
  0.8  = I am the bottleneck desk: the answer cannot move without my tools right now.

Answer with ONE short sentence of reasoning that cites what is already in the conversation, then the number, for example:
<think>one sentence about the evidence so far and what my tools can do next</think>
<bid>0.4</bid>
\end{promptbox}

\textbf{Action Prompts.} 

\begin{promptbox}[mblueL]{mblueD}{Searcher Agent: Act}{box:dr_act_searcher}
[system, appended after the browsing instructions and before the tool schemas]
You are the SEARCHER desk, one of 3 specialists (Searcher/Opener/Finder) that share this browser and this conversation. Exactly one desk acts each turn, chosen by an auction, and this turn is yours. You may ONLY call browser.search, or give the final answer with no tool call. Any other tool call is rejected and the turn is wasted. You decide what to look for: craft queries that decompose the question into findable facts. You cannot open or read pages.

[user, the question and the conversation so far as in the single-agent prompt]
...

\end{promptbox}

\begin{promptbox}[mblueL]{mblueD}{Opener Agent: Act}{box:dr_act_opener}
[system, appended after the browsing instructions and before the tool schemas]
You are the OPENER desk, one of 3 specialists (Searcher/Opener/Finder) that share this browser and this conversation. Exactly one desk acts each turn, chosen by an auction, and this turn is yours. You may ONLY call browser.open, or give the final answer with no tool call. Any other tool call is rejected and the turn is wasted. You open pages: follow link ids from the search results or open URLs directly, and scroll with loc / num_lines to read further. You cannot run searches or find text in a page.

[user, the question and the conversation so far as in the single-agent prompt]
...

\end{promptbox}

\begin{promptbox}[mblueL]{mblueD}{Finder Agent: Act}{box:dr_act_finder}
[system, appended after the browsing instructions and before the tool schemas]
You are the FINDER desk, one of 3 specialists (Searcher/Opener/Finder) that share this browser and this conversation. Exactly one desk acts each turn, chosen by an auction, and this turn is yours. You may ONLY call browser.find, or give the final answer with no tool call. Any other tool call is rejected and the turn is wasted. You search inside the page that is currently open: find a pattern to confirm or rule out a candidate fact. You cannot run web searches or open pages.

[user, the question and the conversation so far as in the single-agent prompt]
...

\end{promptbox}

\subsubsection{SFT Cold Start}
To teach the models the input and output formats and establish basic browsing capability, we perform an SFT cold start in both settings. The demonstrations are the OpenResearcher-Dataset's own browsing trajectories, produced by gpt-oss with the same three tools over the same corpus. We use 568 assistant turns, including 222 for \textsc{Searcher}, 273 for \textsc{Opener} and 73 for \textsc{Finder}. The monolithic agent is fine-tuned on all action turns. Each society agent is fine-tuned on the action turns it owns. SFT for each model is performed on a single H200 GPU.

We fine-tune all models in bf16 using AdamW with a cosine learning rate schedule. Detailed settings are listed in Table~\ref{tab:hyperparams_dr}.

\subsubsection{Post-training}
Our post-training dataset consists of the 5{,}000 OpenResearcher questions different from SFT. Search services run on CPU nodes. Reinforcement learning is performed on four H200 GPUs. Detailed settings are listed in Table~\ref{tab:hyperparams_dr}.

\begin{table*}[h]
\centering
\caption{Hyperparameters for the Deep Research experiment.}
\label{tab:hyperparams_dr}
\begin{tabularx}{\textwidth}{@{}lX@{}}
\toprule
\textbf{Category} & \textbf{Hyperparameter $=$ Value} \\
\midrule
\textbf{SFT cold start} &
epochs $=$ 2; per\_device\_batch $=$ 1; grad\_accum $=$ 16; learning\_rate $=$ $1\!\times\!10^{-5}$; warmup\_ratio $=$ 0.03; weight\_decay $=$ 0; max\_grad\_norm $=$ 1.0. \\
\midrule
\textbf{Post-training} &
learning\_rate $=$ $1\!\times\!10^{-6}$; train\_batch\_size $=$ 64; grad\_clip $=$ 1.0; entropy\_coeff $=$ 0. \\
\midrule
\textbf{Generation} &
temperature $=$ 1.0; top\_p $=$ 1.0; \\
& max\_prompt\_length $=$ 16{,}384; max\_response\_length $=$ 4{,}096 (action), 128 (bid). \\
\midrule
\textbf{Environment} &
decisions\_per\_episode $=$ 64; search\_results $=$ 10;  judge $=$ \texttt{qwen3.5-flash}. \\
\bottomrule
\end{tabularx}
\end{table*}

\section{Case Study}
\label{app:vis}

\subsection{Embodied Crafting}
\label{app:vis_ec}

\begin{figure}[H]
\centering
\includegraphics[width=\linewidth]{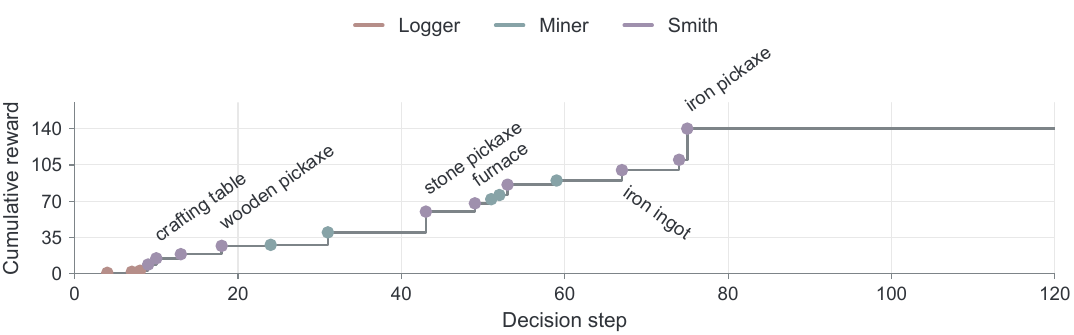}
\caption{\textbf{Embodied Crafting.} Cumulative reward over the 120 decisions of one evaluation episode of the trained population.}
\label{fig:case_ec_reward}
\end{figure}

We walk through one evaluation episode of the trained population in which the society obtains the iron pickaxe. Figure~\ref{fig:case_ec_reward} shows the cumulative reward over the episode; every marker is a decision that received a reward, coloured by the agent that won that decision's auction. The chain unfolds in the order the tech tree requires: \textsc{Logger} gathers logs in the first decisions, \textsc{Smith} converts them into planks, a crafting table, sticks and a wooden pickaxe, \textsc{Miner} then takes over to dig down and mine cobblestone, \textsc{Smith} crafts the stone pickaxe and the furnace, \textsc{Miner} mines iron ore, and \textsc{Smith} smelts the ingots and crafts the iron pickaxe.

\begin{figure}[H]
\centering
\includegraphics[width=\linewidth]{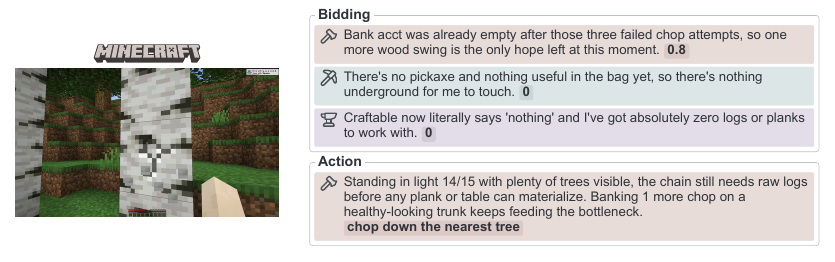}
\caption{\textbf{Decision 3} ($+1$, log). Bids: \textsc{Logger} 0.8, \textsc{Miner} 0, \textsc{Smith} 0; \textsc{Logger} wins and issues ``chop down the nearest tree''.}
\label{fig:case_ec_t003}
\end{figure}

\begin{figure}[H]
\centering
\includegraphics[width=\linewidth]{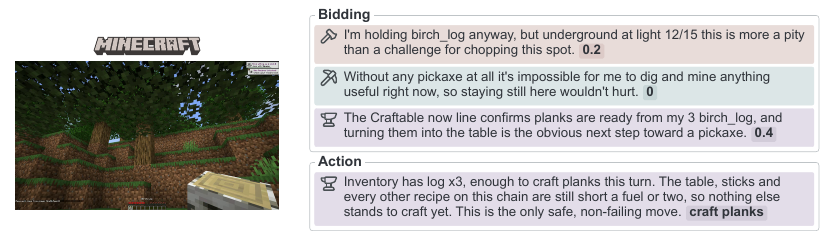}
\caption{\textbf{Decision 8} ($+6$, planks). Bids: \textsc{Logger} 0.2, \textsc{Miner} 0, \textsc{Smith} 0.4; \textsc{Smith} wins and issues ``craft planks''.}
\label{fig:case_ec_t008}
\end{figure}

\begin{figure}[H]
\centering
\includegraphics[width=\linewidth]{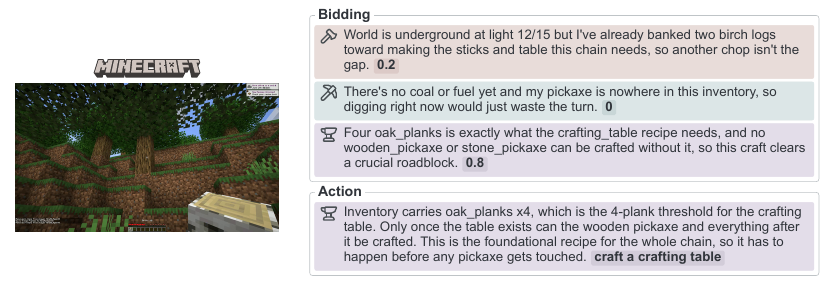}
\caption{\textbf{Decision 9} ($+6$, crafting table). Bids: \textsc{Logger} 0.2, \textsc{Miner} 0, \textsc{Smith} 0.8; \textsc{Smith} wins and issues ``craft a crafting table''.}
\label{fig:case_ec_t009}
\end{figure}

\begin{figure}[H]
\centering
\includegraphics[width=\linewidth]{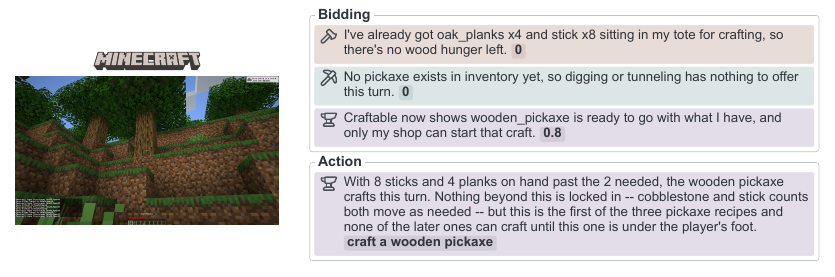}
\caption{\textbf{Decision 17} ($+8$, wooden pickaxe). Bids: \textsc{Logger} 0, \textsc{Miner} 0, \textsc{Smith} 0.8; \textsc{Smith} wins and issues ``craft a wooden pickaxe''.}
\label{fig:case_ec_t017}
\end{figure}

\begin{figure}[H]
\centering
\includegraphics[width=\linewidth]{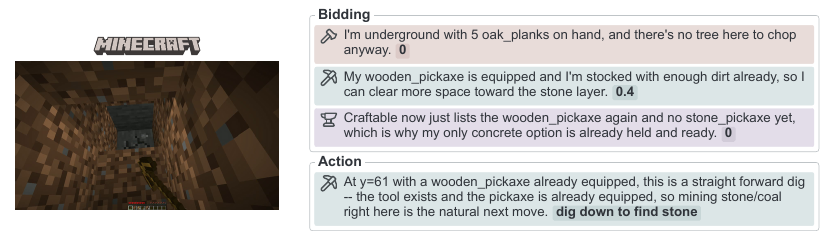}
\caption{\textbf{Decision 30} ($+12$, cobblestone). Bids: \textsc{Logger} 0, \textsc{Miner} 0.4, \textsc{Smith} 0; \textsc{Miner} wins and issues ``dig down to find stone''.}
\label{fig:case_ec_t030}
\end{figure}

\begin{figure}[H]
\centering
\includegraphics[width=\linewidth]{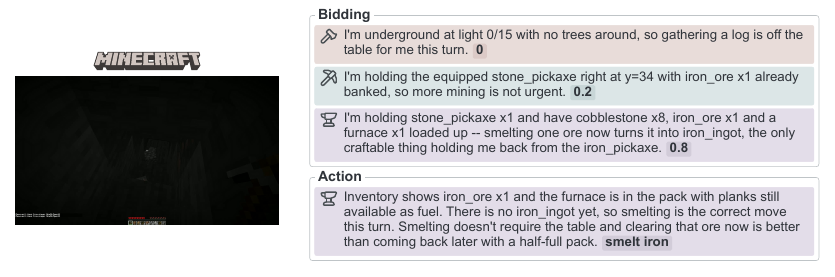}
\caption{\textbf{Decision 52} ($+10$, iron ingot). Bids: \textsc{Logger} 0, \textsc{Miner} 0.2, \textsc{Smith} 0.8; \textsc{Smith} wins and issues ``smelt iron''.}
\label{fig:case_ec_t052}
\end{figure}

\begin{figure}[H]
\centering
\includegraphics[width=\linewidth]{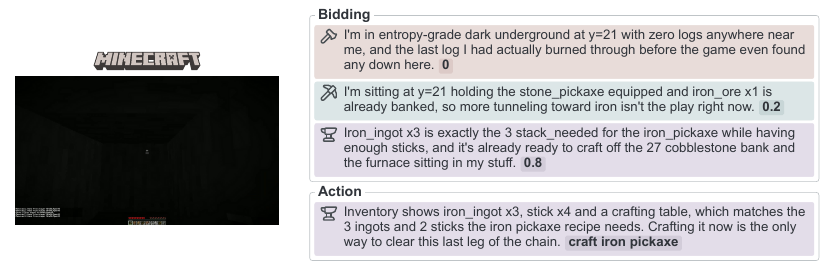}
\caption{\textbf{Decision 74} ($+30$, iron pickaxe). Bids: \textsc{Logger} 0, \textsc{Miner} 0.2, \textsc{Smith} 0.8; \textsc{Smith} wins and issues ``craft iron pickaxe''.}
\label{fig:case_ec_t074}
\end{figure}

Figures~\ref{fig:case_ec_t003}--\ref{fig:case_ec_t074} show example decisions of this episode. Each panel gives the game frame the desks saw at that decision, the three sealed bids with each agent's one-sentence reasoning, and the winner's reasoning and instruction.

\subsection{Robot Navigation}
\label{app:vis_rn}

\begin{figure}[H]
\centering
\includegraphics[width=\linewidth]{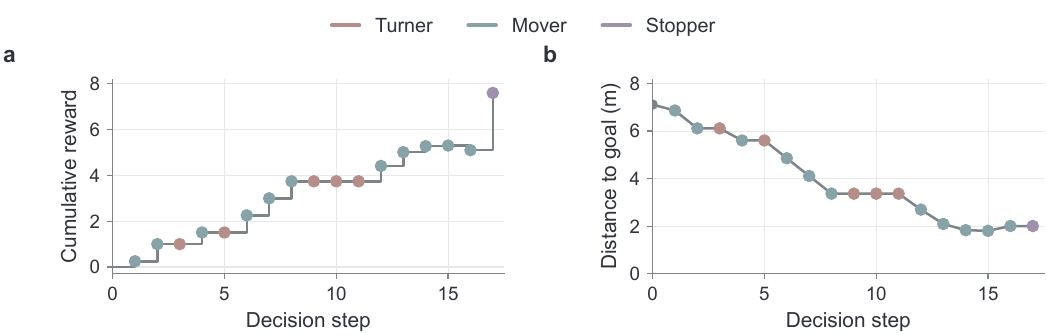}
\caption{\textbf{Robot Navigation.} (a) Cumulative reward and (b) geodesic distance to the goal over the 17 decisions of one episode of the trained population.}
\label{fig:case_rn_reward}
\end{figure}

We walk through one episode of the trained population on a R2R-CE instruction that the society completes: ``Move forward to the doorway on the opposite side of the hall.  Stop in the archway.'' The robot starts 7.1\,m from the goal and stops 2.0\,m from it after 17 decisions, for a return of 7.60. Figure~\ref{fig:case_rn_reward} shows the cumulative reward and the geodesic distance to the goal over the decisions.

\begin{figure}[H]
\centering
\includegraphics[width=\linewidth]{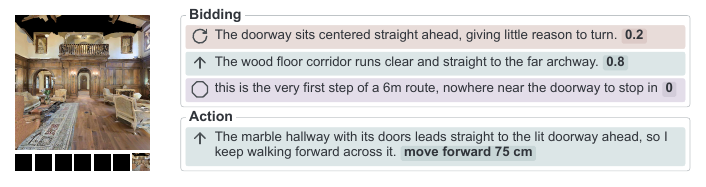}
\caption{\textbf{Decision 2} ($+0.75$, 6.9\,m $\rightarrow$ 6.1\,m). Bids: \textsc{Turner} 0.2, \textsc{Mover} 0.8, \textsc{Stopper} 0; \textsc{Mover} wins and issues ``move forward 75 cm''.}
\label{fig:case_rn_t001}
\end{figure}

\begin{figure}[H]
\centering
\includegraphics[width=\linewidth]{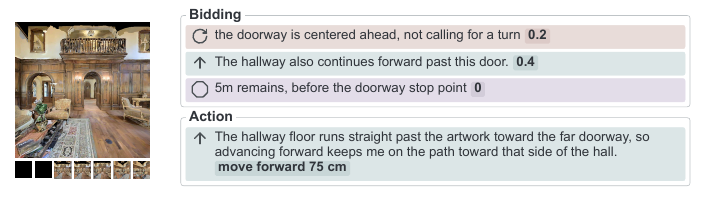}
\caption{\textbf{Decision 6} ($+0.75$, 5.6\,m $\rightarrow$ 4.9\,m). Bids: \textsc{Turner} 0.2, \textsc{Mover} 0.4, \textsc{Stopper} 0; \textsc{Mover} wins and issues ``move forward 75 cm''.}
\label{fig:case_rn_t005}
\end{figure}

\begin{figure}[H]
\centering
\includegraphics[width=\linewidth]{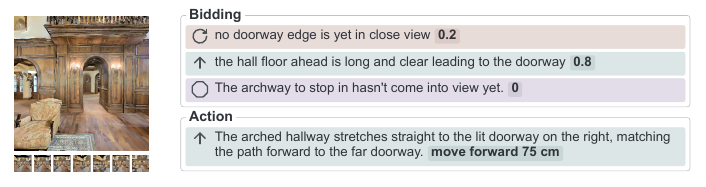}
\caption{\textbf{Decision 8} ($+0.74$, 4.1\,m $\rightarrow$ 3.4\,m). Bids: \textsc{Turner} 0.2, \textsc{Mover} 0.8, \textsc{Stopper} 0; \textsc{Mover} wins and issues ``move forward 75 cm''.}
\label{fig:case_rn_t007}
\end{figure}

\begin{figure}[H]
\centering
\includegraphics[width=\linewidth]{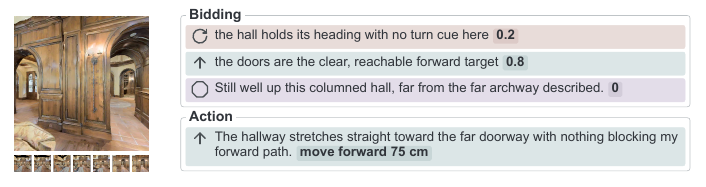}
\caption{\textbf{Decision 13} ($+0.60$, 2.7\,m $\rightarrow$ 2.1\,m). Bids: \textsc{Turner} 0.2, \textsc{Mover} 0.8, \textsc{Stopper} 0; \textsc{Mover} wins and issues ``move forward 75 cm''.}
\label{fig:case_rn_t012}
\end{figure}

\begin{figure}[H]
\centering
\includegraphics[width=\linewidth]{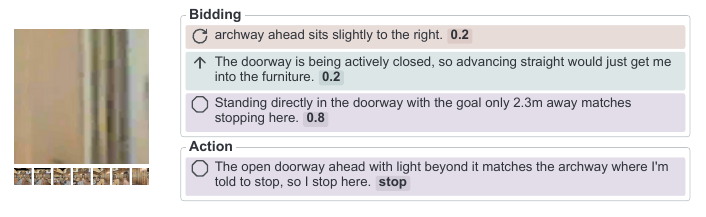}
\caption{\textbf{Decision 17} ($+2.50$, 2.0\,m $\rightarrow$ 2.0\,m, success). Bids: \textsc{Turner} 0.2, \textsc{Mover} 0.2, \textsc{Stopper} 0.8; \textsc{Stopper} wins and issues ``stop''.}
\label{fig:case_rn_t016}
\end{figure}

Figures~\ref{fig:case_rn_t001}--\ref{fig:case_rn_t016} show example decisions of this episode. Each panel gives the eight-frame window the desks saw, the three sealed bids with each desk's one-sentence reasoning, and the winner's reasoning and action.

\subsection{Deep Research}
\label{app:vis_dr}

We walk through one episode of the society on an OpenResearcher question: ``Find the country where the movie `Legend of the Lost' was filmed'' (reference answer: \emph{Libya}). The society answers correctly after 6 decisions. Table~\ref{tab:case_dr_turns} lists every decision. \textsc{Searcher} issues the first query. \textsc{Opener} opens the Wikipedia hit, whose lead paragraph states that location shooting took place near Tripoli, Libya. \textsc{Finder} then searches the open page for the country name and finds further mentions (Leptis Magna, Zliten, Ghadames). Two confirmation lookups follow, and \textsc{Finder} gives the final answer.

\begin{table}[h]
\centering
\caption{\textbf{Deep Research.} All decisions of the episode. Bids are listed as \textsc{Searcher}/\textsc{Opener}/\textsc{Finder}.}
\label{tab:case_dr_turns}
\small
\begin{tabularx}{\textwidth}{@{}clll@{}}
\toprule
\textbf{Decision} & \textbf{Bids (S/O/F)} & \textbf{Winner} & \textbf{Action} \\
\midrule
1 & 0.8 / 0 / 0.2   & \textsc{Searcher} & \texttt{search("Legend of the Lost ... country")} \\
2 & 0.8 / 0.8 / 0.2 & \textsc{Opener}   & \texttt{open(id=0)} $\rightarrow$ Wikipedia: ``\dots near Tripoli, Libya.'' \\
3 & 0.2 / 0.8 / 0.8 & \textsc{Finder}   & \texttt{find("Libya")}  \\
4 & 0.2 / 0.4 / 0.8 & \textsc{Finder}   & \texttt{find("Ghadames")}  \\
5 & 0.2 / 0.4 / 0.8 & \textsc{Finder}   & \texttt{find("Libya")}  \\
6 & 0.2 / 0.2 / 0.8 & \textsc{Finder}   & final answer: \emph{Libya}  \\
\bottomrule
\end{tabularx}
\end{table}

Below we show example decisions. Each box gives the three sealed bids with each agent's one-sentence reasoning, then the winner's full response and the tool output it received.

\begin{promptbox}[mblueL]{mblueD}{Decision 1: \textsc{Searcher} wins}{box:case_dr_t1}
[bids]
Searcher (0.8): <think>Should this 'Legend of the Lost' query help, I'll probably search for instances of movie filming location papers immediately.</think>
Opener   (0):   <think>No search results or cursor pages exist yet, so there is nothing for me to open.</think>
Finder   (0.2): <think>This has been a movie trivia inquiry, so nothing is open yet on this page to spot filmed locations.</think>

[Searcher, act]
<think>
countries for movie legend of the lost filmed
</think>
<tool_call>
{"name": "browser.search", "arguments": {"query": "Legend of the Lost movie filming location country", "topn": 10}}
</tool_call>

[tool_response]
[0] Legend of the Lost movie filming location country (web-search://ts=1790228819)
**viewing lines [0 - 54] of 81**

L4:   * [0\dagger Doc 3902\dagger en.wikipedia.org] Legend of the Lost
L6: From Wikipedia, the free encyclopedia
L8: Legend of the Lost
L9: 1957 movie poster
L10: Directed by      Henry Hathaway
L11: Screenplay by    Ben Hecht
L13:   * [1\dagger Doc 8181659\dagger www.luggagefactory.com] Many of these places are often
  ... (8 more hits, none about this film)
\end{promptbox}

\begin{promptbox}[mblueL]{mblueD}{Decision 2: \textsc{Opener} wins (tie with \textsc{Searcher})}{box:case_dr_t2}
[bids]
Searcher (0.8): <think>The search page already lists legendsrevealed.com but I can search differently to get the location, making a new query better right now.</think>
Opener   (0.8): <think>Once the Wikipedia result is indexed, opening Doc 3902 should reveal geographic filming details.</think>
Finder   (0.2): <think>We have yet to scan the wiki page for mainland location details, but there is no specific location listed yet to find.</think>

[Opener, act]
<think>
wikipedia is great
</think>
<tool_call>
{"name": "browser.open", "arguments": {"cursor": 0, "id": 0}}
</tool_call>

[tool_response]
[1] Doc 3902 (https://en.wikipedia.org/wiki/Legend_of_the_Lost)
**viewing lines [0 - 77] of 121**

L2: Legend of the Lost
L8: Directed by      Henry Hathaway
L13: Starring         John Wayne
L28: Countries        United States
L29:                  Italy
  ...
L34: Legend of the Lost is a 1957 Italian-American adventure film produced and
L35: directed by Henry Hathaway, shot in Technirama and Technicolor by Jack Cardiff,
L36: and starring John Wayne, Sophia Loren, and Rossano Brazzi. The location shooting
L37: for the film took place near Tripoli, Libya.
  ...
\end{promptbox}

\begin{promptbox}[mblueL]{mblueD}{Decision 6: \textsc{Finder} wins and answers}{box:case_dr_t6}
[bids]
Searcher (0.2): <think>This query already opened the page found by Leger of the Lost, so a new search right now would just rerun a find.</think>
Opener   (0.2): <think>The wikipedia page section directly confirming 'Libya' is already open, so opening another search result page would be new context.</think>
Finder   (0.8): <think>The page shows finding results for Libya transitively linking Tripoli, Libya, and Ghadames, so I'll look for the last page reference to confirm.*</think>

[Finder, act]
<think>
Confirmed.
</think>
<final_answer>Libya</final_answer>

[judge]
extracted_final_answer: Libya
reasoning: The extracted final answer "Libya" is an exact match to the provided [correct_answer] "Libya". There are no inconsistencies or ambiguities.
correct: yes
\end{promptbox}

\section{Potential Limitations}
\label{app:lim}

We acknowledge the following limitations:

(1) \ours{} relies on externally evaluated task rewards to train both action and bidding policies. Its effectiveness depends on the quality of these signals. We have not evaluated settings where outcomes are primarily subjective or where reliable feedback is difficult to obtain.

(2) Bidding updates require counterfactual rollouts from the same environment and agent states. Therefore, extending this approach to irreversible real-world interactions remains an open problem.

(3) Despite our efforts to broaden the experimental evaluation, computational constraints limit the scale of our study. We have not explored training swarms composed of multiple models with more than 10B parameters each, or jointly training substantially larger agent populations.

\end{document}